\documentclass[preprint]{aastex}

\slugcomment{}

\shorttitle{Chemical Abundances of the Milky Way Thick Disk and Stellar Halo}
\shortauthors{Ishigaki, Chiba, and Aoki}

\begin{document}


\title{Chemical Abundances of the Milky Way Thick Disk and Stellar Halo I.: Implications of [$\alpha$/Fe] for Star Formation Histories in Their Progenitors}


\author{Miho N. Ishigaki }
\affil{National Astronomical Observatory of Japan}
\affil{2-21-1 Osawa, Mitaka, Tokyo 181-8588, Japan}
\email{ishigaki.miho@nao.ac.jp}

\author{Masashi Chiba }
\affil{Astronomical Institute, Tohoku University}
\affil{Aoba-ku, Sendai 980-8578, Japan}
\email{chiba@astr.tohoku.ac.jp}

\and

\author{Wako Aoki }
\affil{National Astronomical Observatory of Japan}
\affil{2-21-1 Osawa, Mitaka, Tokyo 181-8588, Japan}
\email{aoki.wako@nao.ac.jp}




\begin{abstract}

We present the abundance analysis of 97 nearby metal-poor 
($-3.3<$[Fe/H]$<-0.5$) stars 
having kinematics characteristics of the Milky Way (MW) 
thick disk, inner, and outer stellar halos. 
The high-resolution, high-signal-to-noise optical 
spectra for the sample stars have been 
obtained with the High Dispersion Spectrograph mounted on the 
Subaru Telescope. 
Abundances of Fe, Mg, Si, Ca and Ti have been derived using a
one-dimensional LTE abundance analysis code with Kurucz 
NEWODF model atmospheres. 
By assigning membership of the sample stars to the thick disk, inner or outer 
halo components based on their orbital parameters, 
we examine abundance ratios as a function of [Fe/H] and kinematics
for the three subsamples in wide metallicity
and orbital parameter ranges.
 
We show that, in the metallicity range of $-1.5<$[Fe/H]$\leq -0.5$, 
the thick disk stars show constantly high mean [Mg/Fe] and [Si/Fe] 
ratios with small scatter. 
In contrast, the inner, and the outer halo stars show lower mean 
values of these 
abundance ratios with larger scatter. 
The [Mg/Fe], [Si/Fe] and [Ca/Fe] for the inner and the outer halo stars 
also show weak decreasing trends with [Fe/H] in the range [Fe/H]$>-2$. 
These results favor the scenarios that the MW thick disk formed 
through rapid chemical enrichment primarily through Type II supernovae  
of massive stars, while the stellar halo has formed 
at least in part via accretion of progenitor stellar systems 
having been chemically enriched with different timescales.  

\end{abstract}





\section{Introduction}

The Milky Way (MW) is widely recognized as 
a unique laboratory to test physical mechanisms of galaxy formation 
in the Universe as well as an underlying cosmology through observation 
of resolved stellar populations. 
In the context of galaxy formation theories based on 
the currently standard $\Lambda$CDM model,
the MW is formed through accretions of smaller 
stellar systems like dwarf spheroidal (dSph) galaxies 
\citep[e.g.,][]{diemand07}. 
Numerical simulations have been performed to examine observable 
consequences of this process, 
implementing prescriptions of 
both assembly of dark matter halos and star formation within 
these halos. \citep{bullock05,robertson05,font06a,
  font06b,delucia08,cooper10}.
These studies generally suggest that the accreted galaxies imprint 
distinct substructures in both phase- and chemical abundance-space for 
many Gyrs \citep{font06c,johnston08}, 
demonstrating that the kinematics and chemical abundances are 
an important tracer of the merging history of the Galaxy.

Kinematics and chemical composition of nearby stars 
have provided one of the major constraints on the formation of the 
Galaxy ever since the landmark study of \citet{eggen62}. 
Currently available observational data on the phase-space 
structure of the Galactic stars show signatures of 
such accretion events of dSphs,
as evidenced by spatial distribution and stellar populations 
of halo globular clusters 
\citep{searle78, mackey04, lee07}, streams/overdensities in 
spatial distribution of stars \citep{ibata94,newberg02,
majewski03,juric08}, 
kinematic substructures \citep{helmi99, chiba00, kepley07, dettbarn07,
morrison09,klement09,
starkenburg09, schlaufman09,xue11}, and a metallicity distribution of 
globular clusters \citep{cote00}.
Photometric and spectroscopic studies based on the Sloan Digital Sky Survey 
(SDSS) further advanced our view of the Galaxy.
The three-dimensional map out to the Galactocentric 
distance of $\sim100$ kpc provided by the SDSS 
shows that the substructures and stellar  
streams are ubiquitous, suggesting 
that the Galactic halo has formed, at least in part, 
through satellite accretions 
\citep{belokurov06,juric08,bell08,klement09}.
\citet{carollo10,carollo07} reported that the stellar 
halo is divisible into two components, 
by studying spatial distributions, kinematics and metallicity 
of stars out to a several kpc from the Sun based on the 
SDSS data. Although the net-rotational velocity may be subject to 
some uncertainty in the currently available data 
 (see discussions in \citet{schonrich11,beers12}), 
at least growing number of observations suggest
that the outer region of the Galactic halo 
contains a certain fraction of stars with extreme motions, 
that are difficult to account for with a single component 
halo \citep[e.g.,][]{kinman12}.
These recent results 
highlight complex nature of the process taken place to form the Galactic halo.

To construct a concrete scenario for the MW formation, an 
important next step is examining whether the old Galactic components, 
namely, the thick disk and the stellar halo, are 
chemically divisible into subcomponents and, 
if so, getting insights about masses, luminosities, 
gaseous contents or time of accretions of progenitors 
of each subcomponent. 
Furthermore, it was proposed that 
not only stars accreted from dwarf galaxies, but also those formed 
``in situ'' have played an important role in forming the current 
structure of the Galactic halo \citep[e.g.,][]{zolotov10,font11}. 
In order to make constraints on a relative 
contribution of these processes, detailed elemental abundances 
of stars belonging to each component provide useful signatures 
in addition to their phase-space structures and overall metallicities.

Detailed chemical abundance patterns of individual stars 
have been used as a fossil record of the star forming 
gas cloud at the stellar birthplace, which are likely conserved
longer than the phase-space structure \citep[e.g.,][]{freeman02}.
In particular, $\alpha$-element-to-iron 
abundance ratios ([$\alpha$/Fe]), where $\alpha$ usually stands 
for elements including O, Mg, Si, Ca and Ti, 
are commonly used to characterize
nucleosynthesis and chemical enrichment history 
of a stellar population \citep{tinsley79,matteucci86}.
This is based on the theory that $\alpha$-elements 
are mainly synthesized in massive stars and ejected in Type II 
supernovae (SNe) within a short timescale, while Fe is ejected both in 
Type II and Type Ia SNe, for an extended period of time 
(time scale of Type Ia SNe is estimated to 
distribute from 0.1 Gyr 
to more than 10 Gyr; \citet{kobayashi11}). 
Chemical enrichment of successive  
SNe II/Ia events finally form a characteristic 
trend in [$\alpha$/Fe] with [Fe/H]. The trend 
can depend on various environment of star formation such 
as an efficiency that gas is converted to stars, ability 
that the system can retain their enriched metals or initial 
mass function \citep[IMF; e.g.,][]{lanfranchi03}.
The [$\alpha$/Fe]-[Fe/H] trend is, therefore, considered 
as a useful diagnostic to constrain chemical evolution of 
the structural components of the MW, including  
 the thin disk \citep{edvardsson93}, bulge \citep{fulbright07, 
bensby09, bensby11}, thick disk 
\citep{bensby03,reddy06,bensby07,reddy08}, and 
stellar halo \citep[e.g.,][]{mcwilliam98,cayrel04,lai08,hollek11}.
Similarities and differences between these components were 
also investigated \citep{melendez08,bensby10,reddy06}.
Studies on [$\alpha$/Fe]-[Fe/H] trends further extended to 
the MW dwarf galaxies like relatively bright classical dwarfs
\citep{shetrone01,tolstoy03,shetrone03,venn04,tolstoy09,aoki09a,kirby11,lai11} 
and the recently discovered ultra-faint dwarfs 
\citep{koch08,feltzing09,norris10}. 
In particular, based on a homogeneous analysis of $\alpha$-element abundances 
based on medium resolution spectra, 
\citet{kirby11} examined the trend in [$\alpha$/Fe] with 
[Fe/H] of the individual dwarf MW satellites 
to characterize their star formation history and its relation to
the global properties like stellar/dynamical mass, 
luminosity of these galaxies.
Such trend, therefore, is an important tracer of chemical 
enrichment in possible progenitors of the old components of 
our Galaxy, namely, the thick disk and stellar halo.

Previous studies on chemical abundances of nearby stars 
taking into account information on kinematics  
provide important insights into the hierarchical formation 
of the MW halo \citep{nissen97a,stephens02,fulbright02,gratton03,
roederer08,zhang09,ishigaki10,nissen10,nissen11}.
In particular, \citet{nissen10} (hereafter NS10) found the two distinct populations 
in the halo in terms of [$\alpha$/Fe], namely ``low-$\alpha$'' 
and 'high-$\alpha$' stars, 
by performing a precise differential abundance analysis for 
 the sample stars having similar metallicities and 
atmospheric parameters.

NS10 also found that the ``low-$\alpha$'' stars tend to have 
higher orbital energies. This correspondence between abundances 
and kinematics implies that the low-$\alpha$ stars may 
represent accreted populations. In order to get more 
insights into the relation of these chemical inhomogeneity 
with formation history 
of each structural component of the MW, a homogeneous 
sample that includes stars having characteristic 
kinematics of the thick disk, inner, and outer halo components 
is necessary. Furthermore, the [$\alpha$/Fe] ratios for these three 
components in a wide metallicity range allow us to examine 
possible trend of the [$\alpha$/Fe]  with [Fe/H], which has 
a significant implication to the star formation history in  
progenitors of each Galactic component.

In \citet{zhang09} and \citet{ishigaki10}, we have presented 
an abundance analysis of stars whose orbits reach a distance 
larger than 5 kpc above and below the Galactic plane 
($Z_{\rm max}>5$ kpc) and compared 
their abundances with those of stars having 
smaller $Z_{\rm max}$ from literature. The comparison 
in the abundances was not straightforward because of the possible 
presence of the offset in the results between different studies.
In this paper, we instead adopt the same methods as much as possible 
for all the sample here 
to avoid any related systematics. 
The sample used in this 
study is selected based on their kinematics so that the sample 
stars span a large range in kinematic parameters. In particular, 
the present sample includes stars having characteristic 
kinematics of the thick disk, inner, and outer halo stars classified 
based on the recent estimates of mean of three velocity components 
 and their dispersions from \citet{carollo10}.

Section \ref{sec:sample} describes the selection of our sample 
and their kinematic properties. Then, 
Section \ref{sec:observation} describes observation and 
reduction for the new sample. Section \ref{sec:analysis}, 
provides detailed procedure of the estimation of 
stellar atmospheric parameters and elemental abundances. 
Section  \ref{sec:results} and \ref{sec:discussion} 
show our main results on [$\alpha$/Fe] and discuss 
implications on progenitors of the MW thick disk and 
the stellar halo.

\section{The sample}
\label{sec:sample}

\subsection{ Selection of the sample}
The sample stars for our high-resolution spectroscopy 
were selected from the catalogs of \citet{carney94}, 
\citet{ryan91} and \citet{beers00}.
These catalogs include data on photometry, radial velocity, 
proper motions and metallicity estimates.
 Space motions, Galactic orbital parameters and distances for 
the stars in the catalogs
 were re-derived as described in Section \ref{sec:orbit}.
 
In the present study, the stars having [Fe/H]$<-0.5$ and $V<14.0$ mag 
were selected.
In our previous studies \citep{zhang09,ishigaki10}, we have 
adopted the kinematic criteria of $Z_{\rm max}$, 
greater than 5 kpc, in order to select mainly the stars 
belonging to the outer stellar halo. 
We have supplement this sample with the stars 
having the orbital parameters characteristics of 
the thick disk and the inner stellar halo for the observing 
runs in 2010. The combined sample, which consists of   
 stars having a wide range of metallicity and orbital parameters, 
is analyzed in a homogeneous manner. Therefore, 
this sample is suitable to examine the 
characteristic abundance patterns of the thick disk, inner, and 
outer stellar halos as a function of metallicity and orbital parameters 
with minimal systematic errors.
Observations of the additional targets in 2010
are described in Section \ref{sec:observation}.
  
\subsection{Kinematics of the sample}

\subsubsection{Orbital parameters}
\label{sec:orbit}

 Velocity components in cylindrical coordinates ($V_{R}$, $V_{\phi}$ 
and $V_{Z}$), apo/peri-Galactic distances ($R_{\rm apo}$ and $R_{\rm peri}$) 
and 
maximum distances above/below the Galactic plane ($Z_{\rm max}$) 
were calculated 
as in \citet{chiba00}. In this calculation, proper motions 
were updated based on the revised {\it Hipparcos} for 76 stars \citep{vanleeuwen07}, {\it Tycho-2} for 14 stars \citep{hog00}, UCAC2 for 5 stars 
\citep{zacharias04}, USNO-B for 2 stars \citep{monet03}, 
and LSPM-North catalog for 2 stars \citep{lepine05}. 
For the stars listed in both revised {\it Hipparcos} and 
{\it Tycho-2}, average proper 
motions and their errors are estimated following the method in \citet{martin98}.
Distances have been derived from {\it Hipparcos} parallaxes if their relative 
errors are less than 15 \% (for 20 stars) and photometric ones for 
the remaining stars as in \citet{beers00}.
 Radial velocities were 
updated to the values derived from the high-resolution spectra obtained 
in this study.

\subsubsection{Membership probability }
\label{sec:membership}

Using the orbital parameters obtained above, 
probabilities that each of the sample stars belongs to  
the thick disk, inner halo or outer halo component ($P_{\rm TD}$, 
$P_{\rm IH}$ and $P_{\rm OH}$, respectively) were calculated from the 
following definition: 

\begin{equation}
P_{\rm TD}=f_{1}(Z_{\rm max})\frac{P_{1}}{P},\hspace{1cm}P_{\rm IH}=f_{2}(Z_{\rm max})\frac{P_{2}}{P},\hspace{1cm}P_{\rm OH}=f_{3}(Z_{\rm max})
\frac{P_{3}}{P}
\label{eq:membership}
\end{equation} 
where $P$ and $P_{i}(i=1,2,3)$ are given as

\begin{equation}
P=\Sigma f_{i}P_{i},
\end{equation}

\begin{equation}
P_{i}=K_{i}\exp \left[ -\frac{(V_{R}-<V_{R,i}>)}{2\sigma_{R,i}^{2}}-\frac{(V_{\phi}-<V_{\phi,i}>)}{2\sigma_{\phi,i}^{2}}-\frac{(V_{Z}-<V_{Z,i}>)}{2\sigma_{Z,i}^{2}}\right]
\end{equation}

\begin{equation}
K_{i}=\frac{1}{((2\pi)^{3/2}\sigma_{R,i}\sigma_{\phi,i}\sigma_{Z,i})}.
\end{equation}

In the above definition of the membership probabilities, 
velocity distributions of
the thick disk, inner, and outer halos (subscripts $i=1,2,$ and $3$, 
respectively) are assumed to 
be Gaussian with mean velocities ($<V_{R}>$, $<V_{\phi}>$, and $<V_{Z}>$)  and 
dispersions ($\sigma_{R}$, $\sigma_{\phi}$, and $\sigma_{Z}$) taken from
 the recent estimates based on the SDSS DR7 by \citet{carollo10}. 
We also employ fractional contribution of each component, $f_{i}(Z_{\rm max}), i=1,2,3$ in equation (\ref{eq:membership}), which varies according to a given 
$Z_{\rm max}$ range \citep{carollo10}. 
The adopted values for these parameters from \citet{carollo10} 
are summarized in Table \ref{tab:kinparams}. 
Note that these parameters may be updated by the reanalysis of 
\citet{beers12}, although the basic characteristics for each component
have remained similar and it does not significantly affect 
the membership assignment in the present work. 
Additionally, we impose that $P_{\rm OH}=1.0$ 
and $P_{\rm TD}=P_{\rm IH}=0.0$ for the stars 
having $R_{\rm apo}>15$ kpc, so that the star having an orbit 
with a low $Z_{\rm max}$ but a large $R_{\rm apo}$ is 
classified as an outer halo candidate. In equations (\ref{eq:membership}), 
$P_{\rm TD}$, $P_{\rm IH}$ and $P_{\rm OH}$ are normalized so that they sum up to unity.

\begin{figure}
\includegraphics[angle=270,scale=0.7]{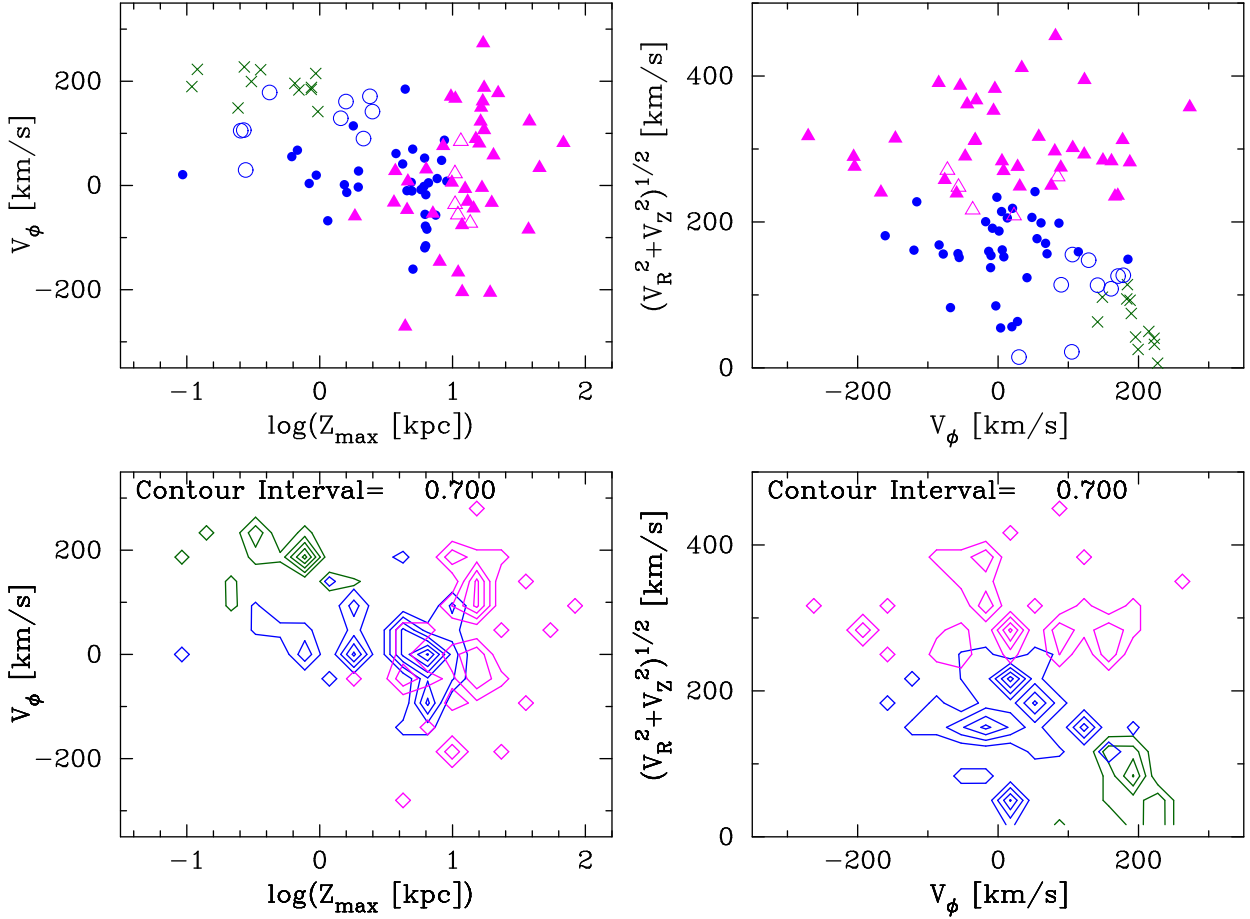}
\caption{Kinematics of the sample stars. The top panels show the 
$\log(Z_{\rm max})-V_{\phi}$ (left) and $V_{\phi}-(V_{R}^{2}+V_{Z}^{2})^{1/2}$ (right) 
diagram.  Crosses, 
filled circles and filled triangles 
indicate the sample stars with $P_{\rm TD}>0.9$ (the thick disk stars), 
$P_{\rm IH}>0.9$ (the inner halo stars) and $P_{\rm OH}>0.9$ 
(the outer halo stars), respectively. Open circles show the 
stars whose kinematics are intermediate between the thick disk 
and the inner halo ( $P_{\rm TD}, P_{\rm IH}\leq 0.9$ and $P_{\rm TD}$, $P_{\rm IH}\geq P_{\rm OH}$ ), 
while open triangles indicate stars 
whose kinematics are 
intermediate between the inner and the outer halo 
($P_{\rm IH}$, $P_{\rm OH}\leq 0.9$ and $P_{\rm IH}$, $P_{\rm OH}\geq P_{\rm TD}$). 
The lower panels show the distributions of the 
 $P_{\rm TD}$ (green), $P_{\rm IH}$(blue) and $P_{\rm OH}$ (magenta), 
in the same diagrams as the top panels.
 Each contour shows sum of the membership probability within a given bin 
of the kinematic parameter.
\label{fig:samplekin}}
\end{figure}

Of the observed sample stars, objects showing broad spectral 
lines, probably due to rapid rotation, or those turn out to 
be metal-rich stars ([Fe/H]$>-0.5$) as a result of our
abundance analysis are excluded 
from the following analysis. As a result, 97 
stars remain.
Top panels of Figure \ref{fig:samplekin} show the  
orbital parameters of the 97 sample stars 
in the $\log Z_{\rm max}$-$V_{\phi}$ (left) and 
$V_{\phi}$-$(V_{R}^2+V_{Z}^2)^{1/2}$ (right) planes. 
Crosses, filled circles and
filled triangles indicate the sample stars 
with $P_{\rm TD}>0.9$, $P_{\rm IH}>0.9$, 
and $P_{\rm OH}>0.9$, respectively. From now on, we will refer to the
three subsamples as the ``thick disk stars'' for $P_{\rm TD}>0.9$, 
``inner halo stars'' for $P_{\rm IH}>0.9$ and ``outer halo stars'' 
for $P_{\rm OH}>0.9$. 
Open circles indicate objects whose kinematics are 
intermediate between the thick disk and the inner halo stars, namely,  
$P_{\rm TD}$, $P_{\rm IH}<0.9$ and $P_{\rm TD}$, $P_{\rm IH}>P_{\rm OH}$.  
Similarly, open triangles indicate objects whose 
kinematics are intermediate
 between the inner and the outer halo stars, namely,$P_{\rm IH}$, $P_{\rm OH}<0.9$ and $P_{\rm IH}$, $P_{\rm OH}>P_{\rm TD}$.

The thick disk stars in the sample have characteristic rotational 
velocities of $V_{\phi}\sim 180$ km s$^{-1}$ and confined to 
$Z_{\rm max}<1$ kpc. Some 
of stars with a small $(V_{R}^2+V_{Z}^2)^{1/2}$ may have
contaminated from the thin disk component. The inner halo stars span a 
wide range in the rotational velocities with an average at 
$V_{\phi}\sim 0.0$ km s$^{-1}$ and are dominant in $Z_{\rm max}<10$ kpc. 
The outer halo stars show a larger dispersion in $V_{\phi}$, some of which 
show extreme prograde or retro-grade rotation.
These stars dominate in $Z_{\rm max}>10$ kpc and in $R_{\rm apo}>10$ kpc, 
which is characterized by high $V_{\phi}$-$(V_{R}^2+V_{Z}^2)^{1/2}$ 
velocities in the Solar neighborhood.

Bottom panels of Figure \ref{fig:samplekin} show 
the distributions of $P_{\rm TD}$ (green), $P_{\rm IH}$ (blue) 
and $P_{\rm OH}$ (magenta) in $\log Z_{\rm max}$-$V_{\phi}$ and
$V_{\phi}$-$(V_{R}^2+V_{Z}^2)^{1/2}$ planes.  As expected from 
its definition, distribution of $P_{\rm TD}$ is peaked at 
$V_{\phi} \sim 180$ km s$^{-1}$ 
and occupies the region of $Z_{\rm max}\leq 1$ kpc. $P_{\rm IH}$ 
spans $-150<V_{\phi}<150$ km s$^{-1}$ and $Z_{\rm max}<10$ kpc. 
$P_{\rm OH}$ spans a wider range in $V_{\phi}$ than $P_{\rm IH}$ and 
is dominant at $Z_{\rm max}>10$ kpc. 

Kinematic parameters for the sample stars and calculated $P_{\rm TD}$, 
$P_{\rm IH}$ and $P_{\rm OH}$ values are summarized in Table \ref{tab:kinematics}.

\section{Observations and data reduction}
\label{sec:observation}

\subsection{Subaru/HDS observations}

The high-resolution spectra were obtained 
in several observing runs during 2003 - 2010 
with the High Dispersion Spectrograph 
\citep[HDS;][]{noguchi02} mounted on the 
Subaru Telescope. The spectra cover 
a wavelength range of 
about 4000-6800 {\AA} except for $\sim 100$ {\AA}
at the gap between the two CCDs at $\sim 5400$ {\AA}. 
 For most of the sample stars, a spectral resolution is
 $R\sim 50000$, while spectra of some sample stars were 
taken with a higher resolution ($R\sim 90000$).  

The data taken before 2010 have already 
been published \citep{zhang09,ishigaki10}. 
Summary of the new observation in 2010 (May 26 and June 18) 
is given in Table \ref{tab:obssummary}.

\subsection{Data reduction}
\begin{figure}
\begin{center}
\includegraphics[height=10.0cm]{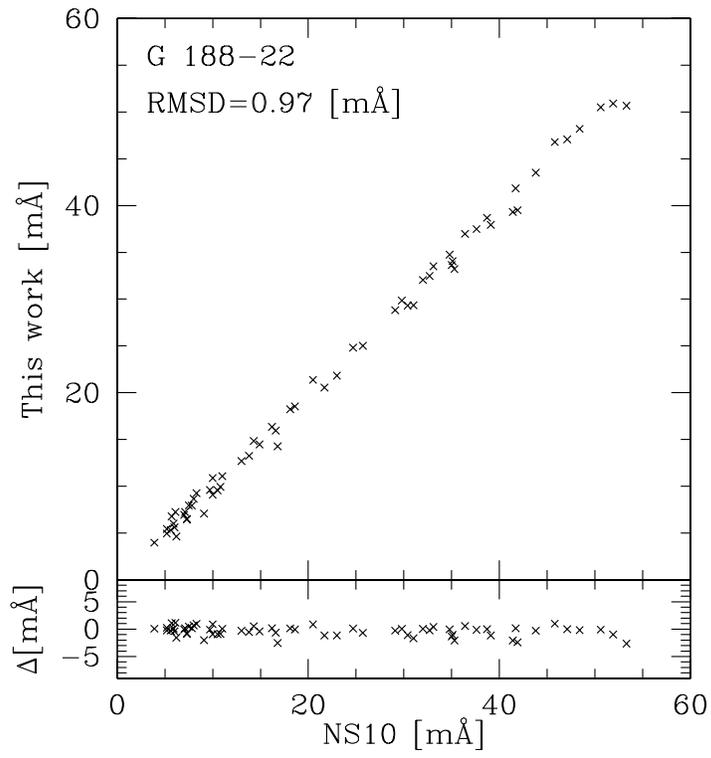}
\end{center}
\caption{Comparison of the measured EWs in this work and those measured in 
\citet{nissen10}}
\label{fig:ews}
\end{figure}

Bias correction, cosmic-ray removal, flat fielding, scattered light 
subtraction, wavelength 
calibration and continuum normalization 
were performed with standard IRAF routines by the same manner 
described in \citet{ishigaki10}. 
The signal-to-noise ratios of the spectra per resolution element, 
after summing up for the spatial pixels, ranges from about 140 to 
about 390 at $\sim 5000$ {\AA} 
(last column of Table \ref{tab:obssummary}).  
Equivalent widths (EW) of \ion{Fe}{1}, \ion{Fe}{2}, \ion{Mg}{1}, \ion{Si}{1}, \ion{Ca}{1}, \ion{Ti}{1} and \ion{Ti}{2} 
lines are measured by fitting Gaussian to each absorption feature. 
The measured EWs for one of the sample stars, 
G 188-22, are in excellent agreement with 
those measured in NS10, with a root-mean-square 
deviation of 0.97 m{\AA} (Figure \ref{fig:ews}). The difference 
in EWs by $<1$ m{\AA} typically affects the derived stellar abundances 
by $<0.03$ dex, which is comparable to or smaller than the random errors 
estimated below. 
The measured EWs are summarized in Table \ref{tab:ews}.

\section{Abundance analysis}
\label{sec:analysis}
The abundance analysis is performed 
using an LTE abundance analysis code 
as in \citet{aoki09b}. 
In this work, effective temperatures ($T_{\rm eff}$) are 
estimated with the Infrared Flux Method (IRFM) as described 
in Section \ref{sec:teff} instead of using excitation 
energies of Fe I lines as adopted in \citet{ishigaki10}. 
This choice has been made to avoid getting unusually low $T_{\rm eff}$ 
and surface gravity ($\log g$) when these are constrained simultaneously 
using the Fe excitation energies and ionization balance, 
likely caused by non-local thermodynamic equilibrium (NLTE) 
effects on Fe abundances 
\citep{bergemann_etal11,mashonkina11}.
As an example, the iterative method adopted in \citet{ishigaki10}
to constrain these parameters  
for CD-24 17504, one of the most metal-poor stars in our sample, 
converges at $T_{\rm eff}=5821$ K and $\log g=3.5$ dex, 
that are lowered by $>300$ K and $>0.8$ dex than those 
reported in literature \citep{aoki09b,nissen07}. Similar problem 
was noticed by \citet{hosford09} and suggested to be 
caused by departure from the assumption of LTE for metal-poor 
stars.  
Therefore, in the present study, we independently estimate 
$T_{\rm eff}$ using $V-K$ colors, while $\log g$ is estimated with
 the Fe I/Fe II ionization balance. Iterations are preformed 
over $\log g$, microturbulent velocity ($\xi$) and [Fe/H] so that these parameters 
are reproduced, while $T_{\rm eff}$ are fixed to their 
photometric estimates.

\subsection{Atomic data}
\label{sec:atomic}

We have selected \ion{Fe}{1} and \ion{Fe}{2} lines from  \citet{ivans06}
and NS10 while their $\log gf$ values are taken mainly from \citet{fuhr06}. 
We restrict the Fe I lines used for the abundance analysis to 
those having a laboratory measurement of the 
$\log gf$ value in the original references
 given in \citet{fuhr06}.
For \ion{Fe}{2} lines, the $\log gf$ values are taken from \citet{melendez09}.
Similarly, the lines of Mg, Si, Ca and Ti are  
selected from both \citet{ivans06} and NS10, while their 
$\log gf$ values are partly replaced based on \citet{aldenius07} and NIST 
for \ion{Mg}{1} lines, \citet{garz73} for \ion{Si}{1} lines, \citet{aldenius09} for \ion{Ca}{1} 
lines and \citet{pickering01} for \ion{Ti}{2} lines. The number of lines used 
is larger than those used in \citet{ishigaki10} for most of the sample stars, 
although the line-to-line scatter is not significantly improved 
by the $\log gf$ updates. 
The adopted $\log gf$ values and their references are given  
in electronic form of Table \ref{tab:ews}.

\subsection{Model atmospheres} 
Model atmospheres of \citet{castelli03} are used in the 
abundance analysis. The model grid covers a range 
in the atmospheric parameter of our sample stars 
($4000<T_{\rm eff}<6900$, $0.0<\log g<5.0$ and $-3.3<$[Fe/H]$<-0.5$). 
For all of our sample stars we have used models with enhanced $\alpha$ 
element abundance of [$\alpha$/Fe]=0.4. Although there are some 
sample stars with lower [$\alpha$/Fe], 
the difference in the derived abundances from those obtained using 
a lower [$\alpha$/Fe] model atmosphere is negligible,
in the precision of the present analysis.
The model atmospheres assume a plane-parallel geometry \citep{castelli03}, 
which may not be a good approximation for giants with low surface 
gravity \citep{heiter06}.  
In order to test whether the low-gravity giants in our sample 
are affected by the geometry effect, we perform the abundance analysis 
for some of these stars using the MARCS model 
atmospheres \citep{gustafsson08}, which 
incorporate spherical geometry for low-gravity stars. 
The results of this analysis are described in Section \ref{sec:marcscomp} .

\subsection{Effective temperature}
\label{sec:teff}

\begin{figure}
\begin{center}
\includegraphics[height=10.0cm]{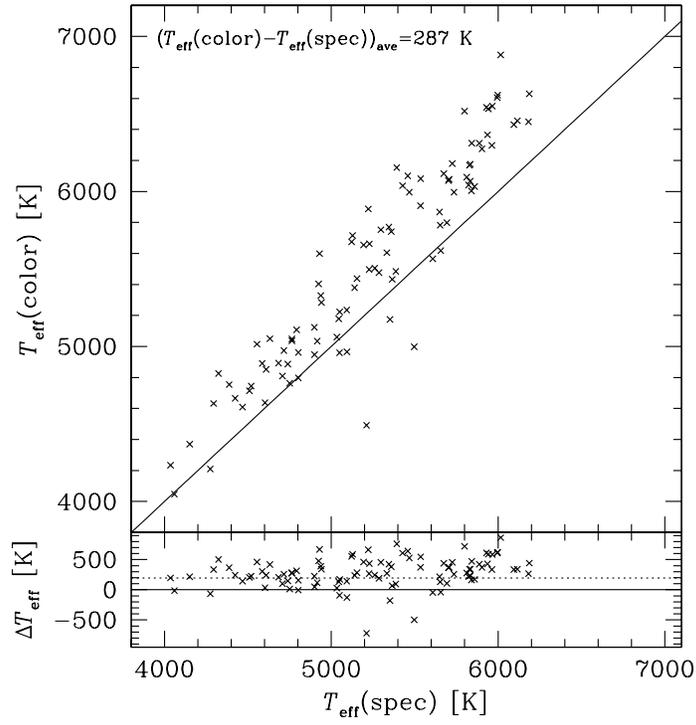}
\end{center}
\caption{Comparison of the color and spectroscopic $T_{\rm eff}$; the top panel shows a plot of $T_{\rm eff}$ estimated from {\it V-K} with that estimated from the abundance-$\chi$ relation of \ion{Fe}{1} lines. The bottom panel plots differences of the two $T_{\rm eff}$ estimates. The mean of the differences is shown as a dotted line.}
\label{fig:teff}
\end{figure}

The $T_{\rm eff}$ is estimated from {\it V-K} colors using 
metallicity-dependent calibrations of \citet{casagrande10} for the sample
stars with $\log g>3.5$ (dwarf and subgiants) and \citet{ramirez05} for 
those with $\log g\leq 3.5$ (giants). 
{\it V} magnitudes and {\it E(B-V)} are taken from 
\citet{carney94}, \citet{ryan91} 
or \citet{beers00}, while {\it E(B-V)} are revised to 
correct for the finite distance to the stars in the method of \citet{beers00}.
{\it K} magnitudes are taken from the Two Micron All Sky 
Survey \citep{cutri03}. 
For comparison, we have estimated the $T_{\rm eff}$ 
using \ion{Fe}{1} lines by minimizing trend in derived abundances 
from individual \ion{Fe}{1} lines with their excitation potentials (EPs)
as has been done in \citet{ishigaki10}. 
The result for the comparison is shown 
in Figure \ref{fig:teff}. The $T_{\rm eff}$ estimated 
from the {\it V-K} colors are 
higher than those from the excitation equilibrium 
by $\Delta T_{\rm eff}=287$ K on average
with a scatter $\sigma=247$ K.  
This difference between photometric and spectroscopic $T_{\rm eff}$ 
is larger than those reported by \citet{casagrande10}  
for a sample of very metal-poor stars, 
$\Delta T_{\rm eff}=177 \pm 33$ K 
or $240 \pm 32$ K  with scatters 
$\sigma=122$ K  or $116$ K, depending on the 
assumed surface gravity. As suggested by \citet{casagrande10} 
the $T_{\rm eff}$ from the excitation equilibrium strongly depend on an
assumed value of the surface gravity. Therefore, the larger scatter 
in this work is likely caused by strong coupling between 
the excitation equilibrium $T_{\rm eff}$ and the surface 
gravity estimated from \ion{Fe}{1}/\ion{Fe}{2} ionization balance.

When the $T_{\rm eff}$ from the ${V-K}$ colors are adopted, 
non-negligible abundance-EP trends, that tend to be negative, appear 
as has been found by \citet{lai08} for their very metal-poor 
sample. Magnitude of the trends for the present sample 
is typically less than 0.1 dex eV$^{-1}$
with a median value of $-0.06$ dex eV$^{-1}$. 
\citet{mashonkina11} examined the abundance-EP trends in a 
LTE and NLTE \ion{Fe}{1} line abundance calculations for 
their sample of metal-poor stars and reported that 
 the negative trends found by LTE analysis are
reduced by the NLTE calculation. 
Therefore, in the assumption of LTE, the negative 
trends may not necessarily be resulted from an
 incorrect $T_{\rm eff}$ scale but may be caused by the 
NLTE effects.

The adopted $T_{\rm eff}$ values are given in Table \ref{tab:stpm_ab}.

\subsection{Surface gravity and microturbulent velocity}
The $\log g$ values are estimated based on \ion{Fe}{1} / \ion{Fe}{2} 
ionization balance as done by \citet{ishigaki10}. 
We have checked the estimated $\log g$ for one of the 
sample star (G 188-22), which has 
relatively good {\it Hipparcos} parallax of $\pi=9.03\pm 1.68$ 
\citep{vanleeuwen07} 
in the method described in \citep{nissen97b}. 
Assuming the $T_{\rm eff}$ and $V_{0}$ magnitudes estimated above 
and a stellar mass of $0.75 M_{\sun}$ for this star, the $\log g$ 
based on the {\it Hipparcos} parallax is 4.38$^{+0.15}_{-0.18}$ dex (errors 
only from the parallax), which is in a good agreement with 4.52 dex, derived 
from the ionization balance. 
For some of the sample stars, the $\log g$ would exceed 5.0 if we attempt to get the 
same [\ion{Fe}{1}/H] and [\ion{Fe}{2}/H]. 
Since $\log g$ values higher than 5.0 are not expected from isochrones \citep[e.g.,][]{demarque04}, 
we simply adopt $\log g=5.0$ for such stars. 
  
The $\xi$ is estimated so that the trend in iron abundances 
estimated from individual \ion{Fe}{1} lines with the EWs 
is minimized as done by \citet{ishigaki10}. 
  
The adopted $\log g$ and $\xi$ values are given in Table \ref{tab:stpm_ab}.

\subsection{Abundances}

The abundances of Mg, Si, Ca and Ti 
are determined from the measured EWs together with
the revised $\log gf$ values as described in Section \ref{sec:atomic}.
Strong lines with reduced EWs ($\log {\rm EW}/\lambda$) greater than 
$-4.7$ (EW$\gtrsim 100$ m{\AA}) are excluded from the analysis, because of 
their larger errors in the EW measurements and the 
saturation effects. 
 
The derived abundances are normalized with the solar values 
from \citet{asplund09} to obtain the [X/H]. The [X/Fe] ratios
are then derived  by normalizing the [X/H] 
with [\ion{Fe}{1}/H] or [\ion{Fe}{2}/H] 
for neutral or ionized species, respectively.

Table \ref{tab:comp_ns10} gives comparisons of the 
derived abundance ratios in 
this work with those from NS10 for nine sample stars, 
that are analyzed in common. 
The [Mg/Fe], [Si/Fe] and [Ca/Fe] are slightly lower 
than those from NS10. However, the mean difference is 
$\lesssim 0.1$ dex, which is comparable to the errors in the present work.  
 The mean differences and standard deviations 
are $(\Delta{\rm [Mg/Fe]}_{\rm TW-NS10},\sigma_{\Delta {\rm Mg}})=(-0.08,0.05)$,
$(\Delta{\rm [Si/Fe]}_{\rm TW-NS10},\sigma_{\Delta {\rm Si}})=(-0.06,0.07)$,
$(\Delta{\rm [Ca/Fe]}_{\rm TW-NS10},\sigma_{\Delta {\rm Ca}})=(-0.10,0.08)$ and
$(\Delta{\rm [Ti/Fe]}_{\rm TW-NS10},\sigma_{\Delta {\rm Ti}})=(0.09,0.03)$, 
indicating that the agreements are fairly good.

\subsection{Errors}

\subsubsection{Scatters in abundances from individual lines}

Random errors of the [X/Fe] ratios are estimated 
by dividing the standard deviation of the abundances 
from individual lines by a square-root of the number of 
lines used to compute the [X/H] values and adding 
the errors in [\ion{Fe}{1}/H] ([\ion{Fe}{2}/H] in the 
case of the ionized species) and those in the Solar abundance \citep{asplund09} 
in quadrature. If only one line is 
available, then the standard 
deviation in the \ion{Fe}{1} abundances, which are typically 0.04 to 0.13 dex,
 is adopted as a measure of 
the random error for that species.

\subsubsection{Errors due to the atmospheric parameters}
Additional errors due to the uncertainties in the adopted 
atmospheric parameters are examined by 
changing the parameters 
by $\pm \sigma_{T_{\rm eff}}$, $\pm \sigma_{\log g}$ and 
$\pm \sigma_{\xi}$. The $\sigma_{T_{\rm eff}}=100$ K is 
adopted for all the sample stars,
although the actual errors in $T_{\rm eff}$ may vary 
from object to object. The
$\sigma_{\log g}$ and $\sigma_{\xi}$ 
are assumed to be 0.3 dex and 0.3 km s$^{-1}$, respectively 
for all sample stars. Table \ref{tab:error} shows the result 
of this exercise for two dwarfs (G24--3, G64--37) and two giants 
(HD 215601, BD--18\arcdeg271), having various metallicities
  in our sample. Typically, 
the deviations due to the change in 
the atmospheric parameters are less than 0.1 dex for  
dwarf stars, regardless of their metallicity. 
In the case of the two giant stars, [\ion{Fe}{1}/H] ratios are sensitive 
to the change in $T_{\rm eff}$, such that $-1.40^{+0.14}_{-0.14}$ dex for HD 215601
and $-2.58^{+0.16}_{-0.17}$ dex for BD--18\arcdeg271. 

We quote the final errors 
as the quadratic sum of these contributions (errors due 
to the line-to-line scatter, $\sigma_{T_{\rm eff}}$, $\sigma_{\log g}$ 
and $\sigma_{\xi}$).  The derived 
abundances and errors are summarized in Table \ref{tab:stpm_ab}.

\subsubsection{Systematic errors due to the model atmospheres} 
\label{sec:marcscomp}

\begin{figure}
\begin{center}
\includegraphics[height=10.0cm]{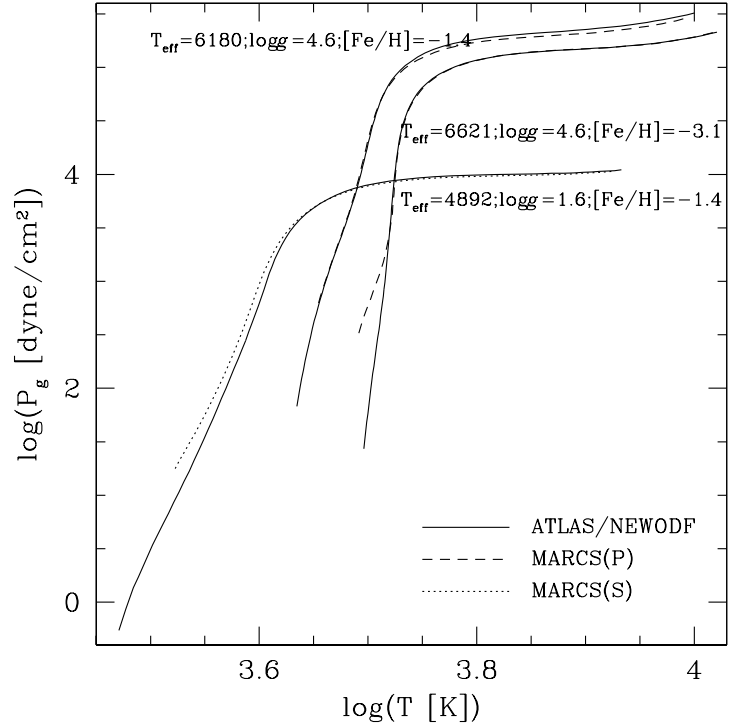}
\end{center}
\caption{Comparison of the Kurucz NEWODF model atmosphere, adopted in this work, with MARCS model atmosphere for three stars in our sample, G 24--3 ($T_{\rm eff}=6180$ K, $\log g=4.6$ dex and [Fe/H]$=-1.4$), HD 215601 ($T_{\rm eff}=4892$ K, $\log g=1.6$ dex and [Fe/H]$=-1.4$) and G 64--37 ($T_{\rm eff}=6621$ K, $\log g=4.6$ dex and [Fe/H]$=-3.1$). Solid, dashed and dotted lines show $T-P_{g}$ relations for the model atmospheres of ATLAS/NEWODF, MARCS/Plain-Parallel and MARCS/Spherical, respectively. The differences in derived abundance ratios are given in the last column of Table \ref{tab:error}.}
\label{fig:opma}
\end{figure}

Figure \ref{fig:opma} shows $T-P_{g}$ relations of 
ATLAS/NEWODF (solid line) and MARCS (dashed 
lines for Plane-Parallel and a dotted line for Spherical) 
model atmospheres for three stars in our sample, G 24--3, 
HD 215601 and G 64--37. The model grids ($T$, $\log P_{g}$) 
have been interpolated to match the adopted atmospheric parameters.
 The differences in the derived abundance ratios are given in 
the last column of Table \ref{tab:error}.

For G 24--3, which is a dwarf star with a metallicity of $-1.4$, 
the models of the ATLAS and MARCS Plane-Parallel are in good agreement. 
Resulting differences in abundance ratios are $\le 0.06$ dex,
which is comparable to the precision of our analysis. 
In contrast, a giant star HD 215601 
having a similar metallicity as G 24--3, the spherical MARCS model is cooler than 
the plane-parallel ATLAS/NEWODF model, at the upper atmospheric layers. 
This difference due to the assumed geometry results in the 
[\ion{Fe}{1}/H] abundance lower by $0.16$ dex when the MARCS 
spherical model is adopted. 
We note that the line calculation in the present work still assumes the
plane-parallel geometry even in the case that 
 the spherical MARCS model is adopted. Therefore, it is possible that 
the deviation listed here is still different from the value 
that would be obtained by a fully consistent spherical analysis 
\citep{heiter06}.
Among the species examined for HD 215601, deviation in the [\ion{Fe}{1}/H] 
ratios is the largest, likely because 
the \ion{Fe}{1} are minor species in the atmosphere of this star. 
In contrast, \ion{Fe}{2}, which is more dominant species,
 is relatively insensitive to the geometry effect. 
In total, the direction and magnitudes of the geometry effect on 
the [\ion{Fe}{1}/H] 
and [\ion{Fe}{2}/H] is in good agreement  with the trend 
obtained by \citet{heiter06} for the solar-metallicity case.  
For other element 
ratios, the geometry effect tends to cancel out and remains to be 
less than 0.1 dex. 
For G 64--37, which is more metal poor than the other two stars, 
the difference in the atmospheric structure is only evident in the 
upper layer in Figure \ref{fig:opma} and resulting differences in the 
abundance ratios are $\le 0.01$ dex, which are almost negligible.

For these three sample stars, the differences in 
the atmospheric structures between the ATLAS and the MARCS model, as can be seen in Figure \ref{fig:opma},
 result in difference in the abundance ratios comparable or much smaller than systematic 
errors from other sources (e.g., $T_{\rm eff}$, $\log g$ or $\xi$). 
As expected, the plane-parallel geometry would not be a very good approximation 
for giant stars as seen in the deviation in [\ion{Fe}{1}/H] by $>0.1$ dex 
when the MARCS 
spherical model is adopted instead of the ATLAS plane-parallel model 
for HD 215601. Such geometry effect is expected to be relatively small
for lines with EW$\lesssim 100$ m{\AA} as mainly used in 
the present work in
the examination by \citet{heiter06} for a wide range of $T_{\rm eff}$ and $\log g$ with solar-metallicity. The effect is much smaller in 
atmospheres with lower metallicity as examined above. 
Furthermore, the effect tends to cancel out in part when the ratio of 
the two abundances of either neutral or ionized species is taken.   
Therefore, although full ranges in atmospheric parameters are not 
examined here, we adopt the ATLAS/NEWODF model atmosphere, 
assuming that the geometry 
effect remains to be small in the [X/Fe] ratios for our 
sample stars in order to keep homogeneity in the analysis method. 
For more high-precision analysis, however, especially for late-type giant 
stars,
fully consistent analysis of spherical geometry is clearly desirable.

\subsection{[X/Fe]-$T_{\rm eff}$ correlation}
\label{sec:abu_teff}

\begin{figure}
\begin{center}
\includegraphics[height=10.0cm]{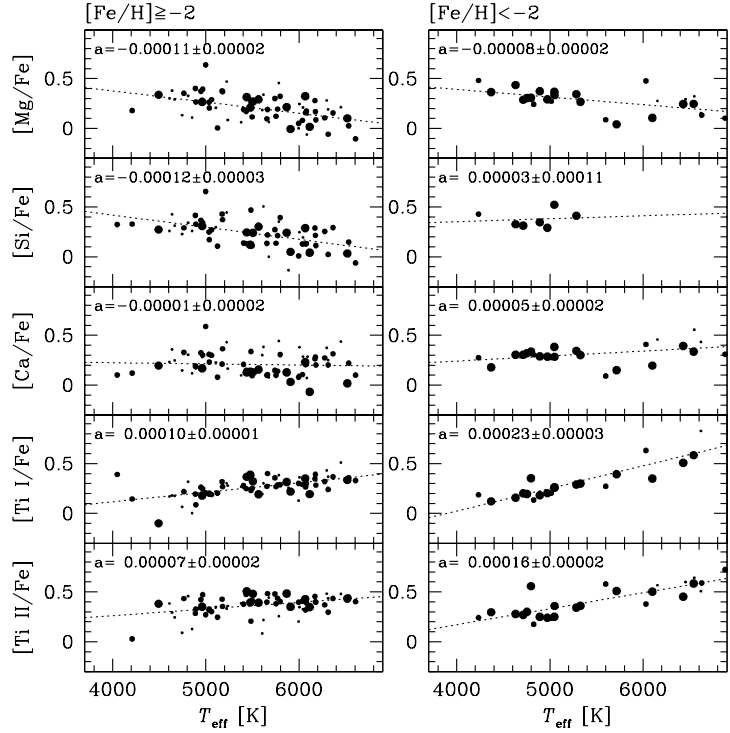}
\end{center}
\caption{[X/Fe] ratios plotted against $T_{\rm eff}$ values for the sample stars with 
the metallicity [Fe/H]$\geq -2$ (left) and [Fe/H]$<-2$ (right). Size of the symbols corresponds to metallicity; for the left (right) panel, small:$-2.0\leq$[Fe/H]$<-1.5$ ([Fe/H]$<-3.0$), medium:$-1.5\leq$[Fe/H]$<-1.0$ ($-3.0\leq$[Fe/H]$<-2.5$), and large:$-1.0\geq$[Fe/H] ($-2.5\leq$[Fe/H]$<-2.0$). Dotted line in 
each panel shows the result of a least square fit to a straight line [X/Fe]$=b+aT_{\rm eff}$. 
The slope $a$ of the fit is indicated in the each panel. 
}
\label{fig:abu_teff}
\end{figure}

Our sample stars have a wide range of $T_{\rm eff}$ from 4000 to 6900 K. 
To examine any systematic differences in derived 
abundances for cool and warm 
stars, Figure \ref{fig:abu_teff} shows the [X/Fe] ratios plotted against the
$T_{\rm eff}$ in the metallicity of [Fe/H]$\geq-2$ (left) 
and [Fe/H]$<-2$ (right). Size of the symbols corresponds to metallicity 
(e.g., larger symbols for higher metallicity).
One may concern that apparent [X/Fe]-$T_{\rm eff}$ correlation 
could arise from possible correlation of [X/Fe] with [Fe/H]. 
This effect might be small because each of the warmer ($T_{\rm eff}>5000$ K) 
and cooler ($T_{\rm eff}<5000$ K) $T_{\rm eff}$ ranges contains 
a certain fraction of 
more metal-rich and more metal-poor stars and is not biased toward/against
 the particular metallicity stars.

In [Fe/H]$<-2.0$, significant positive correlation with a 
slope $\sim 0.1$ dex per 500 K 
can be seen for both [\ion{Ti}{1}/Fe] and [\ion{Ti}{2}/Fe]. 
Calculation of linear correlation coefficients, $r$, and their significance 
level ($P(r)$) for the null hypothesis of zero correlation is 
$P(r)<0.001$\% for the both species.
The values of the slope are similar  
to those reported by \citet{lai08} for their sample of very metal-poor stars.
Such trends may be attributed to NLTE effects on the Ti abundance 
as reported by \citet{bergemann11}. 
 Marginal negative correlation with $T_{\rm eff}$ can be seen for the
[Mg/Fe] and [Si/Fe] ratios in
[Fe/H]$\geq-2.0$, although the slope is $\sim 0.1$ dex per 1000 K, 
which is comparable to the estimated errors of these abundance ratios. 
Care must be taken, however, in examining these abundance 
ratios as a function of [Fe/H] or orbital parameters so that 
the sample is not biased toward/against cooler or warmer stars. 

\begin{figure}
\begin{center}
\includegraphics[height=10.0cm]{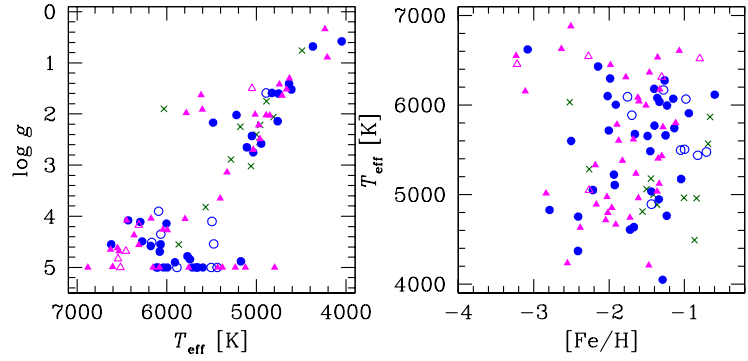}
\end{center}
\caption{
The $\log g$ vs. $T_{\rm eff}$ (left) and $T_{\rm eff}$ vs.
[Fe/H] (right) diagrams for the sample stars. 
Symbols are the same as in the top panels of Figure \ref{fig:samplekin}. 
}
\label{fig:teff_logg_fe_mem}
\end{figure}

To see whether each of the thick disk, inner, and outer halo subsamples, 
defined in Section \ref{sec:membership}, is biased toward/against 
stars with particular atmospheric parameters,  
Figure \ref{fig:teff_logg_fe_mem} shows the $\log g$-$T_{\rm eff}$ (left) 
and $T_{\rm eff}$-[Fe/H] (right) diagrams for the sample stars. 
Symbols are the same as in the top panels of Figure \ref{fig:samplekin}.
It can be seen that the thick disk stars (green crosses) are 
predominantly giant stars while the inner (blue circles) 
and outer halo (magenta triangles) stars 
include larger number of dwarf stars. 
In particular, in a metallicity range of [Fe/H]$>-1.5$,  
the thick disk stars in the sample 
tend to be cooler than the inner and the outer halo stars. 

The difference in typical $T_{\rm eff}$ between 
the thick disk, inner, and outer halo subsample 
might affect comparison in abundances between these 
subsamples. 
The average $T_{\rm eff}$ in the metallicity range of 
[Fe/H]$>-1.5$ is 5108 K for the thick disk, 5608 K for 
the inner halo and the 5676 K for the outer halo. 
According to the observed [Mg/Fe]-$T_{\rm eff}$ slopes in 
Figure \ref{fig:abu_teff}, 
the lower $T_{\rm eff}$ by $500$ K corresponds to 
the [Mg/Fe] by 0.06 dex, although 
the slope could also arise from possible intrinsic abundance 
difference between the thick disk and inner/outer stellar halos.

\section{Results}
\label{sec:results}

\subsection{Distribution of the sample stars in [X/Fe]-[Fe/H] planes}
Left panels of Figure \ref{fig:mg_si_ca} and 
\ref{fig:ti} show the [Mg/Fe], [Si/Fe], [Ca/Fe], [\ion{Ti}{1}/Fe], [\ion{Ti}{2}/Fe] and [Ti/Fe]$=$([\ion{Ti}{1}/Fe]+[\ion{Ti}{2}/Fe])/2 abundance ratios
plotted against the [Fe/H] for 
the thick disk stars ($P_{\rm TD}>0.9$; 
crosses), the inner halo stars ($P_{\rm IH}>0.9$; filled circles) 
and the outer halo stars ($P_{\rm OH}>0.9$; filled triangles). 
Symbols are the same as in Figure \ref{fig:samplekin}.
 Means ($\mu$) and standard deviations 
($\sigma$) of the abundance ratios for each of the thick disk 
and inner/outer halo stars 
within a given metallicity interval are summarized in Table \ref{tab:mean_dev}.
One object showing unusually high [X/Fe] ($0.5-0.6$ dex) values is 
excluded from this calculation and following discussion. 
Since this object shows a very large slope ($\sim 0.1$ dex eV$^{-1}$) in the 
Fe I abundances versus EPs (See Section \ref{sec:teff}), 
the high [X/Fe] values are probably resulted from 
the incorrect $T_{\rm eff}$ adopted in the abundance estimate.

\subsubsection{[Fe/H]$>-1.5$}
\label{subsec:alpha_highmetal}

In the metallicity range [Fe/H]$>-1.5$, the thick disk stars 
in the present sample show enhanced [Mg/Fe] and
 [Si/Fe] ratios at the means larger than 0.3 dex 
with very small scatters ($\sigma \le 0.07$ dex; Table \ref{tab:mean_dev}). The constantly 
high values of these abundance ratios for the thick disk 
stars with [Fe/H]$<-0.5$ are consistent with those reported in  
previous studies \citep[e.g.,][]{bensby03}. 
In contrast, the inner and the outer halo stars in the 
present sample show the [Mg/Fe] and [Si/Fe] ratios $\lesssim 0.2$ dex,  
which is lower than the thick disk stars. 
The scatter for these stars is much 
larger ($\sim 0.12-0.13$ dex) than that of the thick disk stars, 
although contribution from the measurement errors is 
not ruled out. 
A Kolmogorov-Smirnov test yields 
that the probability for a null hypothesis 
that the [X/Fe] ratios of the thick disk stars
are drawn from the same distribution as those of 
the inner halo stars is 2 \% for [Mg/Fe] ratios and 
3 \% for [Si/Fe] ratios. Similarly, the null hypothesis 
for the thick disk and the outer halo is rejected
at the level of $<1$ \% for both [Mg/Fe] and [Si/Fe] ratios. 

To examine whether the observed difference in the [Mg/Fe] 
and [Si/Fe] ratios between the thick disk and inner/outer 
halo stars is caused by the difference in typical $T_{\rm eff}$ 
between these subsamples as described in Section \ref{sec:abu_teff}, 
a comparison was performed using a limited sample of giant stars  
($T_{\rm eff}\le 5500$ and $\log g\le 3.5$) in a metallicity range of 
[Fe/H]$>-1.5$, 
which includes six thick disk, four inner halo and two outer halo stars. 
As a result, the differences in the mean [Mg/Fe] and [Si/Fe] values 
between the thick disk and the outer halo remained to be 
$\sim 0.1$ dex, while the difference in the mean values 
between the thick disk and 
the inner halo vanishes. 
Although this comparison needed to be confirmed with a 
larger sample, this result may indicate the importance of 
understanding systematic errors in derived abundances 
between dwarfs and giants.
In the highest metallicity of [Fe/H]$>-1.0$, differences in the 
[Mg/Fe] and [Si/Fe] for the four thick disk stars and the two inner 
halo stars are significantly large ($\sim 0.3$ dex), which cannot 
be explained by the difference in the typical $T_{\rm eff}$ alone.

In the precise differential analysis of NS10, the sample 
stars in this metallicity range are separated into the 'high-$\alpha$' 
and the ``low-$\alpha$'' stars, defined based on the [Mg/Fe] ratios.
The presence of the chemically distinct components 
in halo stars could not be 
evaluated in the present work because of the lower internal 
precision as described above.  
The inner halo stars kinematically resemble the 
high-$\alpha$ stars in NS10. Some of the inner halo 
stars in the present sample have high [Mg/Fe] similar to 
the thick disk stars, while the two most metal-rich 
inner halo stars show low [Mg/Fe], [Si/Fe] and [Ca/Fe]. 
The outer halo stars, that kinematically resemble 
the low-$\alpha$ stars in NS10, generally show 
lower [Mg/Fe] in their highest metallicity range 
and the overlap with the thick disk stars is small.

The [Ca/Fe] ratios for the 
thick disk, inner, and outer halo stars show largely 
overlapping distributions. In particular, for the thick disk stars, 
the mean [Ca/Fe] ratio is lower than the [Mg/Fe] and [Si/Fe] 
ratios by $\sim 0.1$ dex (Table \ref{tab:mean_dev}). Such a low 
[Ca/Fe] relative to [Mg/Fe] seen in the thick disk stars, where 
the mean [Ca/Mg] ratio less than the solar value, is not seen in the inner and 
outer halo stars, where the mean [Ca/Mg] is $\geq$0.00.

For all of the thick disk, inner, and outer halo subsamples, 
the [\ion{Ti}{2}/Fe] ratios 
are $\gtrsim 0.1$ dex larger than the [\ion{Ti}{1}/Fe] ratios. 
Such an effect is also reported 
by previous studies \citep[e.g.,][]{lai08}. This may indicate
that an ionization balance assumed in the LTE 
analysis is not valid for Ti \citep{bergemann11}.

\subsubsection{[Fe/H]$<-1.5$}
 
In the metallicity range of [Fe/H]$<-1.5$, the thick disk stars 
are rare and the halo stars dominate.
The thick disk stars in this metallicity range, again, show higher [Mg/Fe] 
ratios ($>0.3$ dex) than the inner/outer halo 
stars with a small scatter. 
Both of the inner and the outer halo stars 
show a wider range in these element ratios from the near-solar 
value to [X/Fe]$\sim 0.5$ than seen in the thick disk stars. 
The scatter in the abundance ratios is 
similar to that of the higher metallicity for all of the three 
subsamples, except for extremely metal-poor stars ([Fe/H]$<-3.0$).

The [Mg/Fe], [Ca/Fe], [\ion{Ti}{1}/Fe] 
and [\ion{Ti}{2}/Fe] values 
of the sample stars in this metallicity range 
generally agree with previous studies 
by \citet{cayrel04}, \citet{lai08} and 
\citet{bonifacio09}, 
which are shown in Figures \ref{fig:mg_si_ca} and \ref{fig:ti} with 
dotted, dash-dotted 
and dashed lines, respectively. The [Si/Fe] ratios are 
slightly higher in this study. 
One of the reasons for this discrepancy could be the difference in the 
\ion{Si}{1} lines used: these studies mainly use the \ion{Si}{1} line 
at 3905.53 {\AA} while 
the present study uses redder lines. In fact, if we use the same 
EWs and $\log gf$ values 
as those used in \citet{lai08} for their one sample star, BD$+$03 740, 
our abundance analysis 
adopting the same atmospheric parameter results in [Si/Fe]$=0.12$, 
which is in good agreement with the value of [Si/Fe]$=0.07$ 
from \citet{lai08} within the quoted error.

\subsubsection{[X/Fe]-[Fe/H] trend}
\label{subsubsec:alpha_fe}

A trend in the [X/Fe] with [Fe/H] is frequently interpreted as a tracer 
of chemical evolution of a self-enriched stellar system \citep{tinsley79,matteucci86,gilmore91}. 
It is particularly 
interesting whether the abundance ratios for the thick disk, 
inner, and the outer halo stars show different 
trends, which could be evidence of different 
star formation history of their progenitors.  
In order to examine the trend for the present sample, 
the right panel of Figures \ref{fig:mg_si_ca} and \ref{fig:ti}
plot the weighted means of the abundance ratios
within a given [Fe/H] interval for each of the subsamples. 
The error bars correspond to the error in the weighted 
means of these abundance ratios. 

In [Fe/H]$>-2.5$, the [Mg/Fe], [Si/Fe] and [Ca/Fe] ratios for 
the inner and the outer halo stars slightly decrease with [Fe/H]. 
 The thick disk stars also show hints of decreases at the highest 
metallicity bin.
This apparent offset, however, could 
be caused by contamination of thin disk stars that are 
known to have lower [$\alpha$/Fe] \citep{lee11}. According to our 
criteria (Section \ref{sec:membership}),  stars 
having disk-like kinematics are all classified as thick disk stars, 
although thin disk stars may present in the range [Fe/H]$>-0.5$.

The [\ion{Ti}{1}/Fe] and [\ion{Ti}{2}/Fe] ratios do not show 
such a decreasing trend but stay constant in [Fe/H]$>-2.5$ 
for all of the three subsamples. 
The absence of the [Ti/Fe]-[Fe/H] trend may result from 
production of Ti in Type Ia SNe together with Fe, 
in contrast to other $\alpha$ elements 
like O or Mg, that are predominantly enriched
via Type II SNe.

In [Fe/H]$<-2.5$, the [X/Fe]-[Fe/H] trends are not seen 
for the three subsamples within the error bars. 
The absence of the trend is expected 
if the chemical enrichment predominantly occurred through 
Type II SNe \citep{kobayashi06}. We note that in this low 
metallicity, the outer halo stars having 
a large retro-grade orbit ($V_{\phi}<-100$ km s$^{-1}$) 
have similar abundance ratios as those of the typical 
inner halo stars with similar metallicity 
on average.

\subsection{Correlation in the [X/Fe] ratios with kinematics}

Figures \ref{fig:alpha_kin} shows the [X/Fe] ratios of the 
thick disk, inner, and outer halo stars 
plotted against the orbital parameters, 
$V_{\phi}$, $\log Z_{\rm max}$, $\log R_{\rm apo}$ and eccentricity 
($e$). Symbols are the same as in Figure \ref{fig:mg_si_ca}. 
Solid and dashed 
lines connect means and means $\pm$ standard deviations, respectively, 
within a given interval of each orbital parameter.
Left and right panels in each diagram show the plots for 
$-1.5<$[Fe/H]$\leq-0.5$ and 
$-2.5<$[Fe/H]$\leq-1.5$, respectively. This separation is made 
so that any [X/Fe]-[Fe/H] correlation would not produce 
apparent [X/Fe] correlation with 
the orbital parameters. For example, because fraction of 
the outer halo stars increases at lower metallicity, 
the possible increase in the [Mg/Fe] with the decreasing 
[Fe/H] would result in high [Mg/Fe] for such outer halo stars having 
large $R_{\rm apo}$ or $Z_{\rm max}$.

In the [X/Fe]-$V_{\phi}$ plot (top left), finite 
correlation is seen in the increasing 
[Mg/Fe] with $V_{\phi}$ at $-1.5\leq$[Fe/H]$<-0.5$. 
In this metallicity range, 
 probability for null correlation is less than 0.1 \% according to 
the calculated linear correlation coefficient. 
This can partly be explained by 
the difference in [Mg/Fe] between the thick disk and 
the halo stars; the former have high $V_{\phi}\sim 200$ km s$^{-1}$ 
while the latter span
a wide range in $V_{\phi}$ as mentioned in Section 
\ref{subsec:alpha_highmetal}. 
If the thick disk stars are excluded from the calculation of 
the correlation coefficient, the probability for the 
null correlation is increased to 4 \%.

In the [X/Fe]-$\log(Z_{\rm max})$ plot (top right), 
the correlation is not significant for these abundance ratios.  
The [Mg/Fe] ratios are slightly lower in $Z_{\rm max}>1$ kpc 
in $-1.5<$[Fe/H]$\leq-0.5$, 
probably because of the absence of the thick disk stars, 
that have high [Mg/Fe], at high $Z_{\rm max}$.

In the [X/Fe]-$\log(R_{\rm apo})$ plot (bottom left), 
the apparent correlation is seen only for the [\ion{Ti}{2}/Fe]
in $-2.5<$[Fe/H]$\leq-1.5$. 
A negative correlation in the 
[Mg/Fe] versus $\log(R_{\rm apo})$ also cannot be ruled out; 
the probability for the null correlation is 11 \%. 
\citet{stephens02} suggested the 
decreasing trend of [$\alpha$/Fe]
(=[(Mg+Si+Ca+Ti)/Fe]) with $R_{\rm apo}$ based on their 
sample of halo stars covering a similar metallicity range as the 
present study, although the observed trend is rather small 
(0.1 dex per decade).  
Since their study as well as the present study use bright local sample, the 
number of the outer halo stars is still not large enough 
to conclude on this subject.

 In the [X/Fe]-$e$ plot (bottom right), a decreasing trend 
of the [Mg/Fe] with $e$ at $-1.5<$[Fe/H]$\leq -0.5$ can be seen. 
This trend is significant at the level that the probability 
for the null correlation is less than 0.1 \%. The probability 
is remained to be significant after the thick disk stars 
are excluded ($<3$ \%). 
This result is 
in qualitative agreement with that of \citet{schuster11}, 
who show that low-$\alpha$ stars dominate 
at the large maximum eccentricity, $e_{\rm max}\geq 0.85$.

\subsection{The membership probabilities in the [X/Fe]-[Fe/H] diagram}
\label{sec:memprob}

In previous sections, we have simply defined the thick disk, inner, and outer
halo stars as being $P_{\rm TD}>0.9$, $P_{\rm IH}>0.9$ and $P_{\rm OH}>0.9$, respectively.
Although these cuts are useful to select potential candidates 
of members of the each Galactic component, they remove  
the sample stars with intermediate kinematics from 
the interpretation of the [X/Fe]-[Fe/H] diagnostics. 
The difference and similarities in the [X/Fe] between the 
three subsamples have only been discussed based on the means 
and the standard deviations within a given metallicity range for these 
limited sample stars (Table \ref{tab:mean_dev}). 
To maximize the use of the [X/Fe]-[Fe/H] diagrams, 
taking into account all of the sample stars with proper weights, 
we examine the distribution of the 
$P_{\rm TD}$, $P_{\rm IH}$, and $P_{\rm OH}$ 
on the [X/Fe]-[Fe/H] plane.   

\begin{figure}
\begin{center}
\includegraphics[height=12.0cm,angle=270]{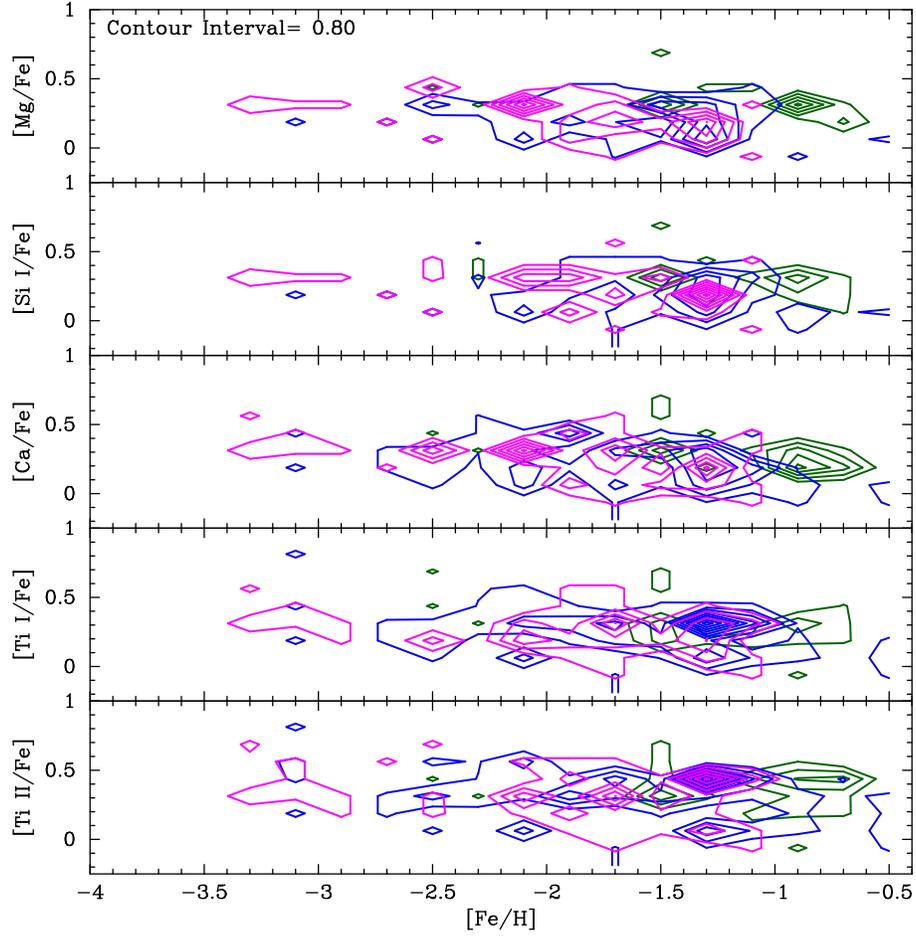}
\caption{ Distribution for $P_{\rm TD}$ (green), $P_{\rm IH}$ (blue) and 
$P_{\rm OH}$ (magenta) in [X/Fe]-[Fe/H] planes.
Each contour shows sum of the membership probability within a given [X/Fe] and [Fe/H] bin.}
\label{fig:alpha_contour}
\end{center}
\end{figure}

Figure \ref{fig:alpha_contour} shows distribution 
of $P_{\rm TD}$ (green), $P_{\rm IH}$ (blue) and 
$P_{\rm OH}$ (magenta) in the [X/Fe]-[Fe/H] planes.  
Each contour shows sum of the membership probabilities 
within a given [X/Fe] and [Fe/H] bin.
This figure illustrates difference and similarity in 
distribution for the thick disk, inner, and outer halo components,
in the [X/Fe]-[Fe/H] planes. 
We note that there may be a selection bias in [Fe/H], since 
only stars with [Fe/H]$<-0.5$ are included in our sample.
Therefore, the distribution of the [Fe/H] is not representative 
of a true underlying distribution of the thick disk, 
inner, and outer halo populations. We therefore, 
restrict our discussion on the difference in the [X/Fe] 
for the three subsamples in a given metallicity range.

As implied from the previous sections, the [Mg/Fe] 
and [Si/Fe] show similar trends with [Fe/H] for each of the $P_{\rm TD}$, 
$P_{\rm IH}$ and $P_{\rm OH}$. 
The membership probability for the thick disk ($P_{\rm TD}$)
 shows a narrow distribution in the [Mg/Fe] and [Si/Fe] 
peaked at $\sim 0.3$ in a range $-1.5<$[Fe/H]$< -0.5$. 
In contrast, the membership probability for the 
inner halo ($P_{\rm IH}$) shows significantly 
broader [Mg/Fe] and [Si/Fe] distributions, that are 
largely overlap with 
the distributions for the $P_{\rm TD}$. 
The membership probability for the outer halo 
($P_{\rm OH}$) shows overlapping distribution 
with those of the $P_{\rm IH}$, but peaked 
at relatively low [Mg/Fe] and [Si/Fe] with 
smaller dispersion than the $P_{\rm IH}$.
It can also be seen that the peak in the [Mg/Fe] 
and [Si/Fe] for the $P_{\rm OH}$ gradually shifts 
from $\sim 0.4$ in [Fe/H]$<-2$ to $\sim 0.2$ in [Fe/H]$>-1.5$. 
The distribution for the $P_{\rm IH}$ and $P_{\rm OH}$ 
is indistinguishable in the lower metallicity range 
([Fe/H]$<-2.0$) 

 In the [Ca/Fe]-[Fe/H] plot, the 
distributions for $P_{\rm TD}$,$P_{\rm IH}$ and $P_{\rm OH}$ 
largely overlap with each other and clear distinctions of the 
peak position in [Ca/Fe] between the three probabilities 
are not seen. 
This behavior is similar in the [\ion{Ti}{1}/Fe] and 
[\ion{Ti}{2}/Fe]-[Fe/H] diagram except that the 
 $P_{\rm IH}$ and $P_{\rm OH}$ appear to 
show double peaks at [Fe/H]$\sim -1.3$. 

In total, if all of the three subsamples are considered, 
distribution in the [X/Fe]-[Fe/H] plane 
is roughly peaked at constant [Mg/Fe]
in a wide metallicity range ($-3.5<$[Fe/H]$<0.5$). 
However, when the thick disk, inner, and outer halos are separately
 considered as in Figure \ref{fig:alpha_contour}, 
differences in distribution are seen for 
[Mg/Fe] and [Si/Fe] ratios.

\section{Discussion}
\label{sec:discussion}

We summarize the main results of the present study as follows.

\begin{itemize}
\item The thick disk stars show relatively high [Mg/Fe] and [Si/Fe] 
ratios compared to 
the inner and outer halo stars for [Fe/H]$>-1.5$. 
\item The inner halo stars show a mean [Mg/Fe] and [Si/Fe] ratios lower 
than the thick disk stars with larger scatter.
\item The outer halo stars show similarly low mean [Mg/Fe] and [Si/Fe] 
ratios as the inner halo stars, which is lower than 
those of the thick disk stars. The outer halo stars (and the part of 
the inner halo stars) also show 
a hint of a decreasing trend in 
[Mg/Fe], [Si/Fe] and [Ca/Fe] with [Fe/H] in [Fe/H]$>-2.5$. 
\item The three subsamples show largely overlapping distribution 
in [Ca/Fe], [\ion{Ti}{1}/Fe] and [\ion{Ti}{2}/Fe]. 
\item Correlation of the [Mg/Fe] ratios with the orbital 
eccentricity $e$ is seen in $-1.5\leq$[Fe/H]$<-0.5$ at a significant level.
For other orbital parameters ($V_{\phi}$, $Z_{\rm max}$, $R_{\rm apo}$), 
significant correlation with [X/Fe] is not clearly seen. 
\end{itemize} 

In the following subsections, we first compare the results 
for the [Mg/Fe], [Si/Fe] and [Ca/Fe] ratios
with those obtained for stars belonging to the 
MW dwarf satellite galaxies. Finally, we discuss 
the implications of the present result for the formation of the MW thick 
disk and stellar halo.

\subsection{Comparison with the chemical abundances of the MW dwarf satellites}
\label{sec:dsph}

\citet{tolstoy09} review recent data on the 
abundances for the MW dwarf satellites including
Fornax, Sculptor, Sagittarius and Carina dSphs. 
In [Fe/H]$>-1.5$, the thick disk, inner, and outer halo stars 
in the present sample show higher [Mg/Fe] than 
those seen in the stars in the dSphs.
For example, the Sculptor dSph shows a near solar average 
value of the [Mg/Fe] ratio 
at [Fe/H]$\sim -1.5$ \citep{tolstoy09}, while the inner and the 
outer halo stars 
at this metallicity show the average [Mg/Fe] of 
$\sim 0.1-0.2$ (Table \ref{tab:mean_dev}). The difference becomes 
more substantial at higher metallicity, where the [Mg/Fe] 
decreases with metallicity in Sculptor, reaching the sub-solar 
mean [Mg/Fe] ratio at [Fe/H]$>-1.0$. 
As can be seen in Figure \ref{fig:alpha_kin}, the sample stars 
having either extreme retrograde orbits, 
high $Z_{\rm max}$ or $R_{\rm apo}$, which are the most likely 
candidates of the accreted stars, do not 
show the [X/Fe] ratios significantly lower than the solar values. 
This difference in abundance ratios between the field halo and dSph stars 
implies that the halo progenitors 
had different chemical enrichment history 
from surviving dSphs 
in terms of star formation rate, galactic wind efficiency or duration 
and frequency of major star formation episodes that drive chemical 
enrichment of these systems. 
In the case of Sculptor, \citet{kirby11} show that 
 the [$\alpha$/Fe]-
[Fe/H] relation for their medium resolution 
spectroscopic sample is well reproduced by 
a model with a low star formation rate, low initial gas mass and 
 the $\sim1$ Gyr duration of star formation which may have 
started more than 10 Gyr ago. This scenario suggests that 
the chemical enrichment of the system proceeds slowly 
so that the overall metallicity of the system remained low 
when Type Ia SNe start to enrich the system with Fe \citep{tolstoy09}. 
The inner and outer halo stars show modest decreasing trends in 
these abundance ratios likely starting at [Fe/H]$> -2.0$. 
The trends, however, are much shallower 
than those seen in Sculptor. 
This result may suggest that the possible
progenitors of the inner and outer halo stars 
have stopped forming stars before the enrichment from 
Type Ia SNe became significant.

In [Fe/H]$<-2.0$, the number of stars studied for individual 
abundances is smaller, probably because of the relative scarcity of 
very metal-poor stars in the well-known classical dSphs. 
Some of the extremely metal-poor stars in the Sextans dSphs show 
near-solar [Mg/Fe] ratios \citep{aoki09a}, that are lower 
than the inner and the outer halo stars in the present sample. 
On the other hand, recently discovered 
ultra-faint dSphs generally show super-solar values of the 
[Mg/Fe] similar to the inner/outer halo stars \citep{tolstoy09}.  
Given that these observed dSphs show a wide range of [Mg/Fe] ratios 
from near-solar to $\sim 1.0$, the scatter in the [Mg/Fe] in 
both the inner and outer halo 
stars are relatively small ($< 0.15$ dex). 
A larger sample in this low metallicity range is desirable 
to characterize the abundance ratios and their scatter 
in the MW halo in comparison with the dSphs.

Different behavior of [Ca/Fe] and [Ti/Fe] 
ratios from those of [Mg/Fe] and [Si/Fe] were previously noted 
for the stars in the MW dwarf satellite galaxies 
\citep[e.g.,][]{venn04}. \citet{letarte10} show that the trend of the 
lower [Ca/Fe] and [\ion{Ti}{1}/Fe] than [Mg/Fe] or [Si/Fe]
for red giant stars is seen in Fornax dSph. They suggest 
that this discrepancy come from either different 
nucleosynthetic origins of Ca and Ti from those of 
Mg and Si or significant dependence of Type Ia SNe 
Ca and Ti yields on metallicity. The latter possibility
is unlikely because if the Type Ia SNe have contributed significantly 
to the chemical evolution of the thick disk, the [Mg/Fe] 
ratios would decrease as the [Fe/H] increases, which is 
inconsistent with the observed high [Mg/Fe] ratios for the 
thick disk stars in the present sample. 
If the former is mainly responsible for the 
low [Ca/Fe] and [\ion{Ti}{1}/Fe] ratios, the thick disk stars would be 
formed in the progenitors that have been enriched more with Mg and Si 
from hydrostatic C and O burning than with Ca and Ti 
from explosive nucleosynthesis in SNe.

\subsection{Implication for the formation of the MW old components}

The constantly high [Mg/Fe] and [Si/Fe] ratios of the thick disk stars, 
unlike the observed dSphs, suggest that these are predominantly 
enriched with Type II SNe 
\citep{kobayashi06}. This suggests that the initial star formation 
in the progenitor of the thick disk stars was high enough to 
enrich the system to [Fe/H]$>-1$ and short enough to complete
 before Type Ia SNe produce 
significant Fe. The low [Ca/Mg] ratios observed for the 
thick disk stars are in qualitative 
agreement with those expected from the Type II SNe yields 
integrated over progenitor 
stellar masses with Salpeter IMF in \citet{tsujimoto95}, supporting 
enrichment in the thick disk progenitors has occurred
predominantly via Type II SNe. 
Thick disk stars in the solar neighborhood 
are also known to be old (age $\leq 12$ Gyr)  suggesting that the 
star formation took place at the early stage of the 
MW formation \citep{fuhrmann11}.

\begin{figure}
\begin{tabular}{cc}
\begin{minipage}{0.5\hsize}
\includegraphics[height=8.0cm]{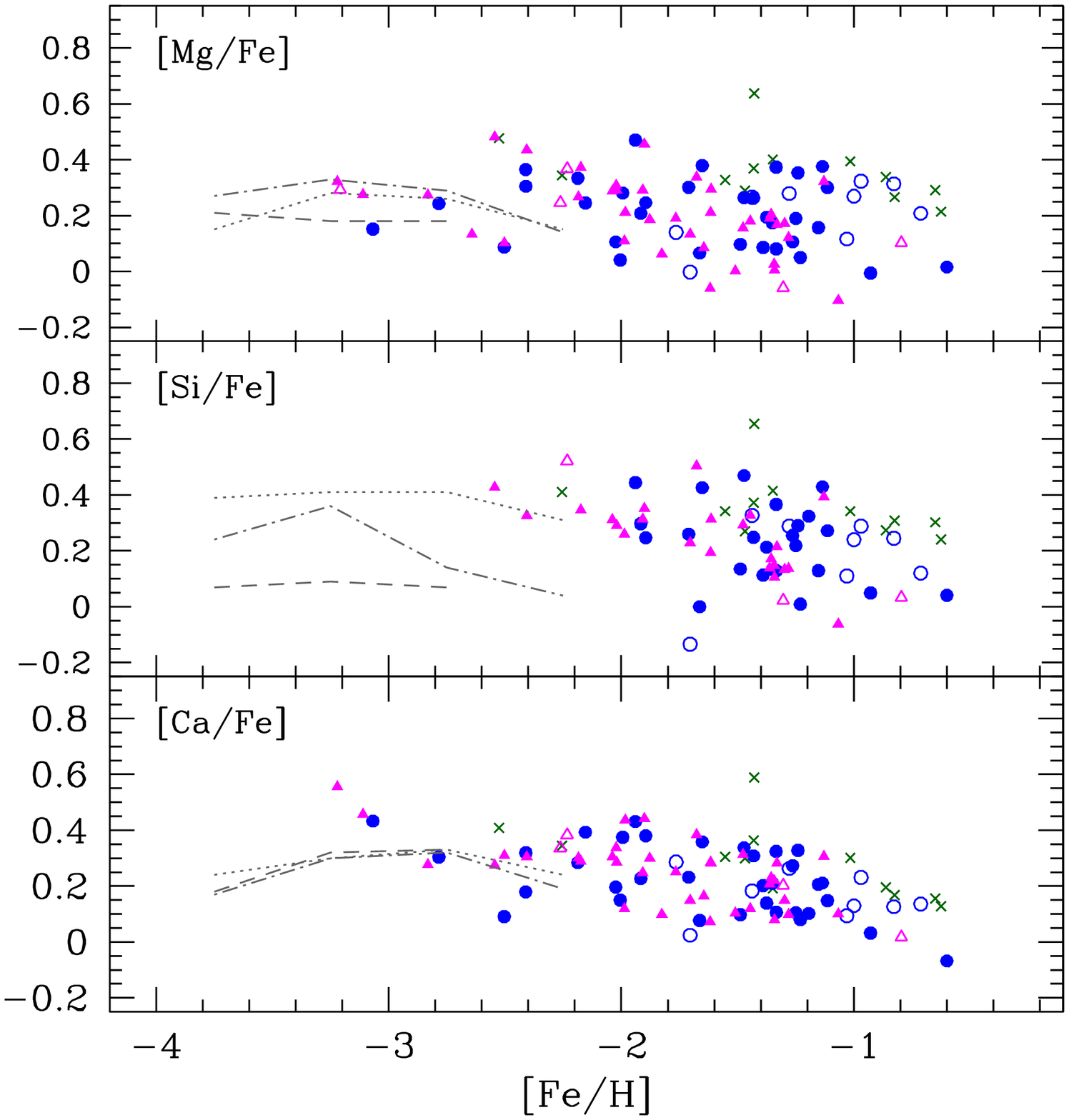}
\end{minipage}
\begin{minipage}{0.5\hsize}
\includegraphics[height=8.0cm]{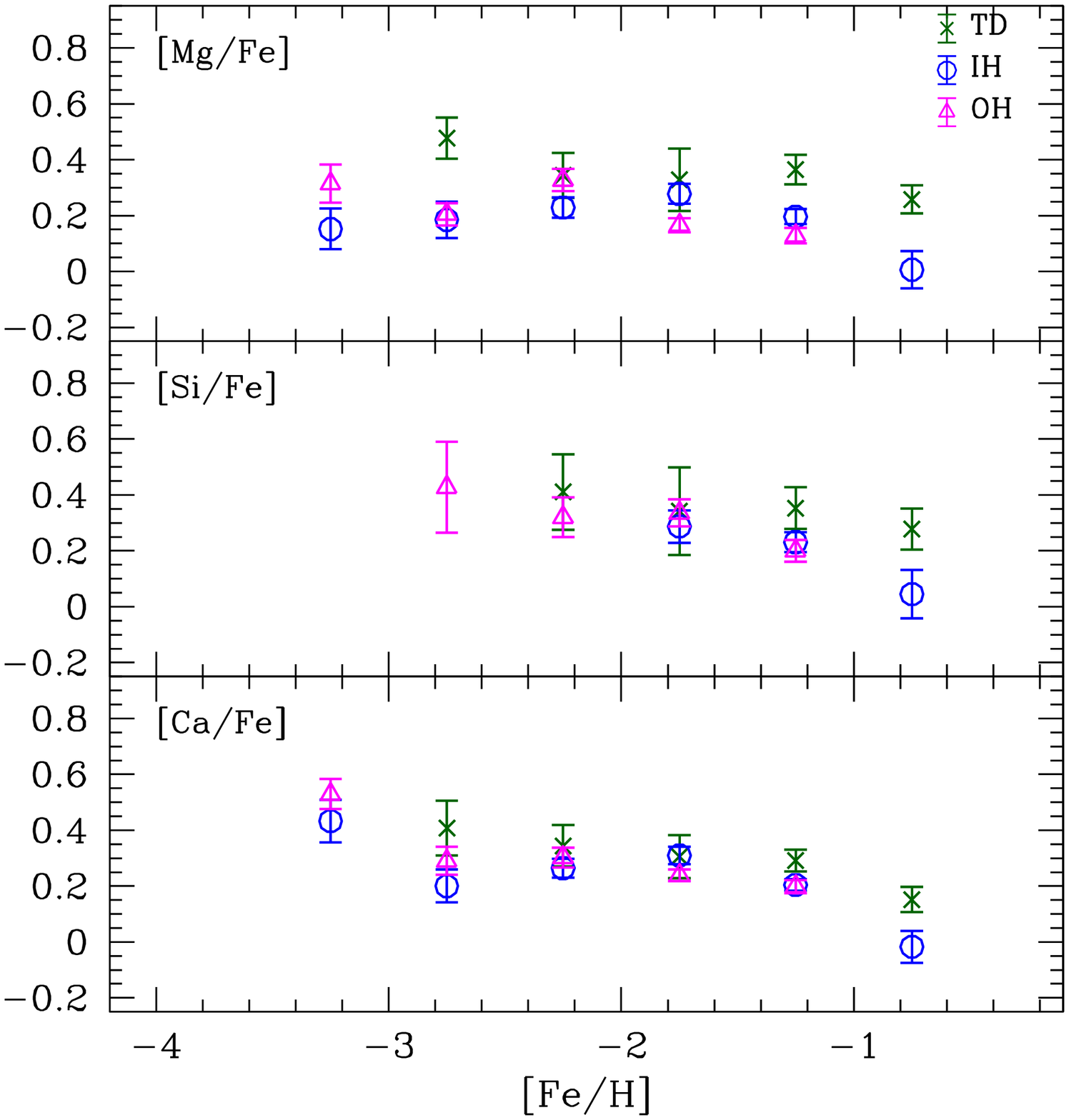}
\end{minipage}
\end{tabular}
\caption{{\it left}: $\alpha$-elements (Mg, Si, Ca) to iron abundance ratios as a function of [Fe/H]. Symbols are the same as in the top panels of Figure \ref{fig:samplekin}. Mean values of the abundance ratios 
within a given metallicity interval obtained by \citet{cayrel04}, \citet{lai08} and \citet{bonifacio09} 
are connected with dotted, dash-dotted and dashed lines, respectively.
 {\it Right}: Weighted mean of the [Mg/Fe], [Si/Fe] and [Ca/Fe] in each [Fe/H] interval for the sample stars with $P_{\rm TD}>0.9$ (crosses), $P_{\rm IH}$ (circles) and $P_{\rm OH}>0.9$ (triangles). }
\label{fig:mg_si_ca}
\end{figure}

\begin{figure}
\begin{tabular}{cc}
\begin{minipage}{0.5\hsize}
\includegraphics[height=8.0cm]{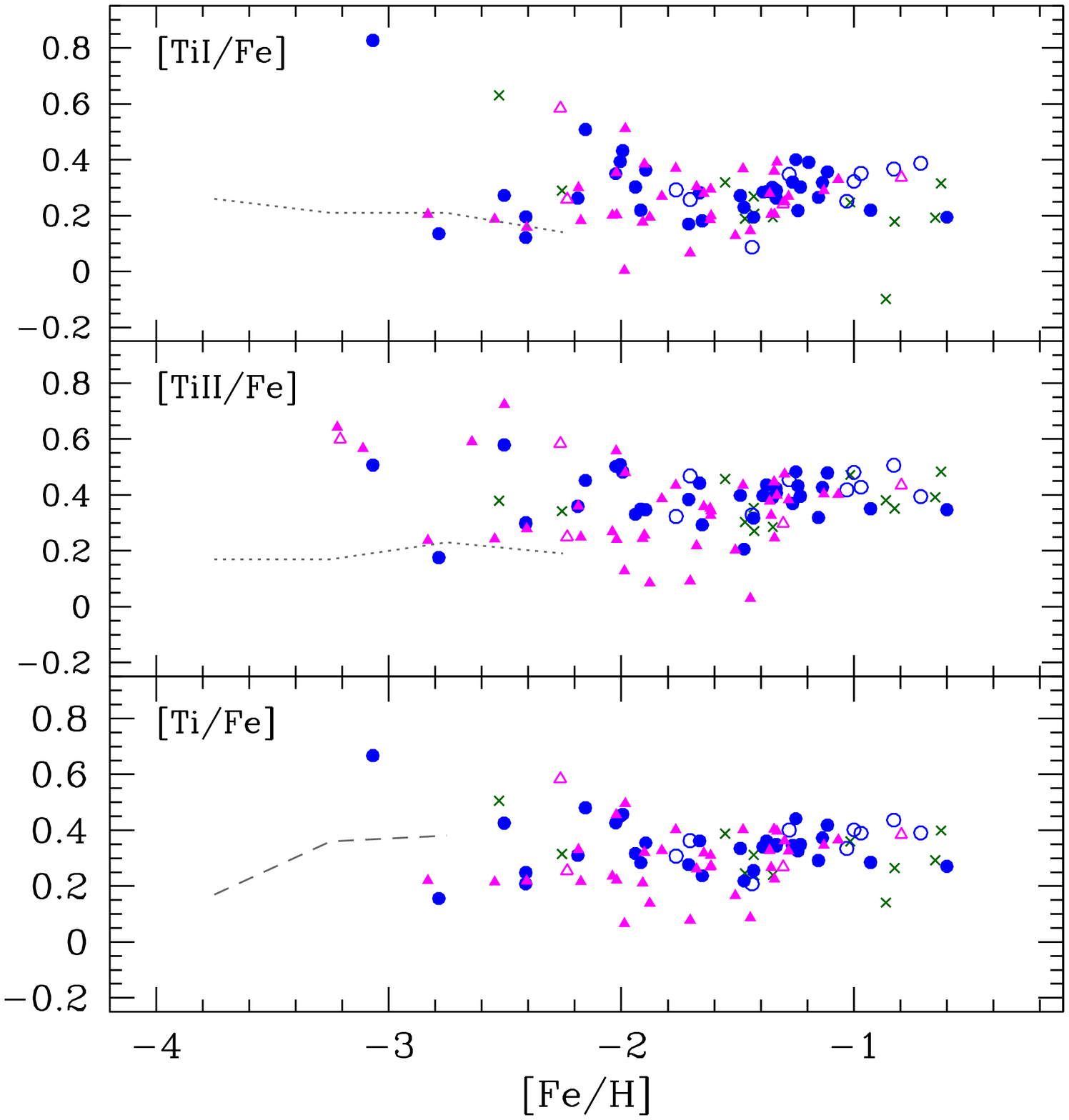}
\end{minipage}
\begin{minipage}{0.5\hsize}
\includegraphics[height=8.0cm]{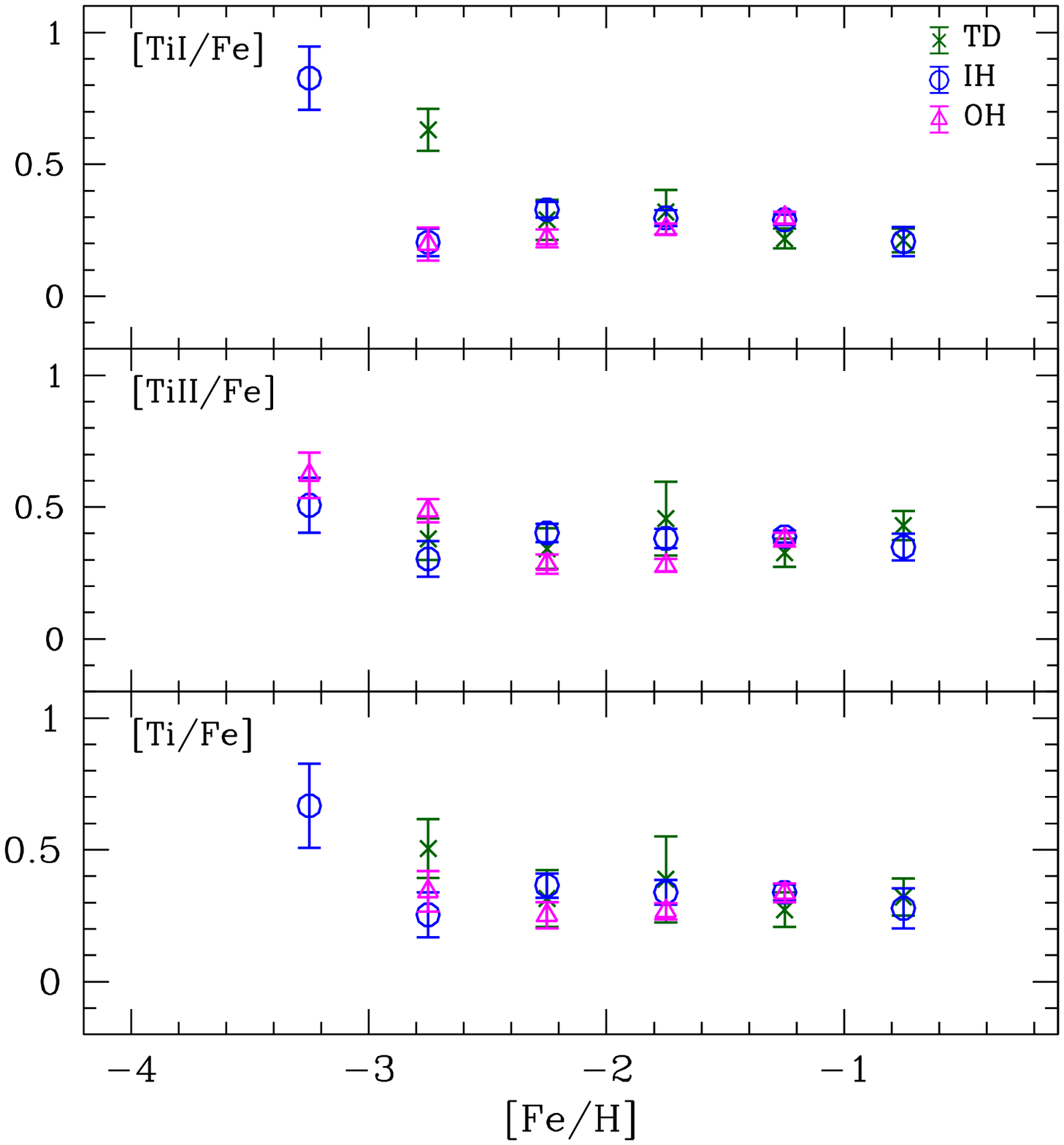}
\end{minipage}
\end{tabular}
\caption{Same as Figure \ref{fig:mg_si_ca} but for [\ion{Ti}{1}/Fe], [\ion{Ti}{2}/Fe] and [Ti/Fe] $=$([\ion{Ti}{1}/Fe]+[\ion{Ti}{2}/Fe]) / 2.}
\label{fig:ti}
\end{figure}

Several mechanisms are proposed for the origin of the 
halo stars, in the context of the hierarchical galaxy formation scenario.   
First, the halo stars could be formed through dissipative 
collapse of gaseous material on to the central region 
of the Galaxy, often referred to as ``in situ'' stars 
\citep{zolotov10,font11}. 
This could be achieved either via rapid collapse of primordial 
gas on to the Galactic dark matter halo 
or through early major mergers of gas-rich galaxies. 
In such a system, high star formation rate is triggered by 
shocks in the interstellar medium (ISM) and, as a result, metal enrichment 
proceeds primarily through Type II SNe with a short time-scale comparable to 
the age of massive stars ($<10^{6}-10^{7}$ yr). 
Thus, this process would yield stars with high [$\alpha$/Fe]. 
In the present study, such high-[$\alpha$/Fe] stars are found 
both in the thick disk and the inner halo subsamples, as seen in the 
[Mg/Fe] and [Si/Fe]-[Fe/H] plots in the left panel of 
Figure \ref{fig:mg_si_ca}.  This result suggests that a sizable 
fraction of the inner halo stars were formed out of gas 
enriched rapidly via Type II SNe before Type Ia SNe became 
a significant contributor to metals in the ISM, 
similar to the thick disk stars.

\begin{figure}
\begin{tabular}{cc}
\begin{minipage}{0.5\hsize}
\includegraphics[height=8cm]{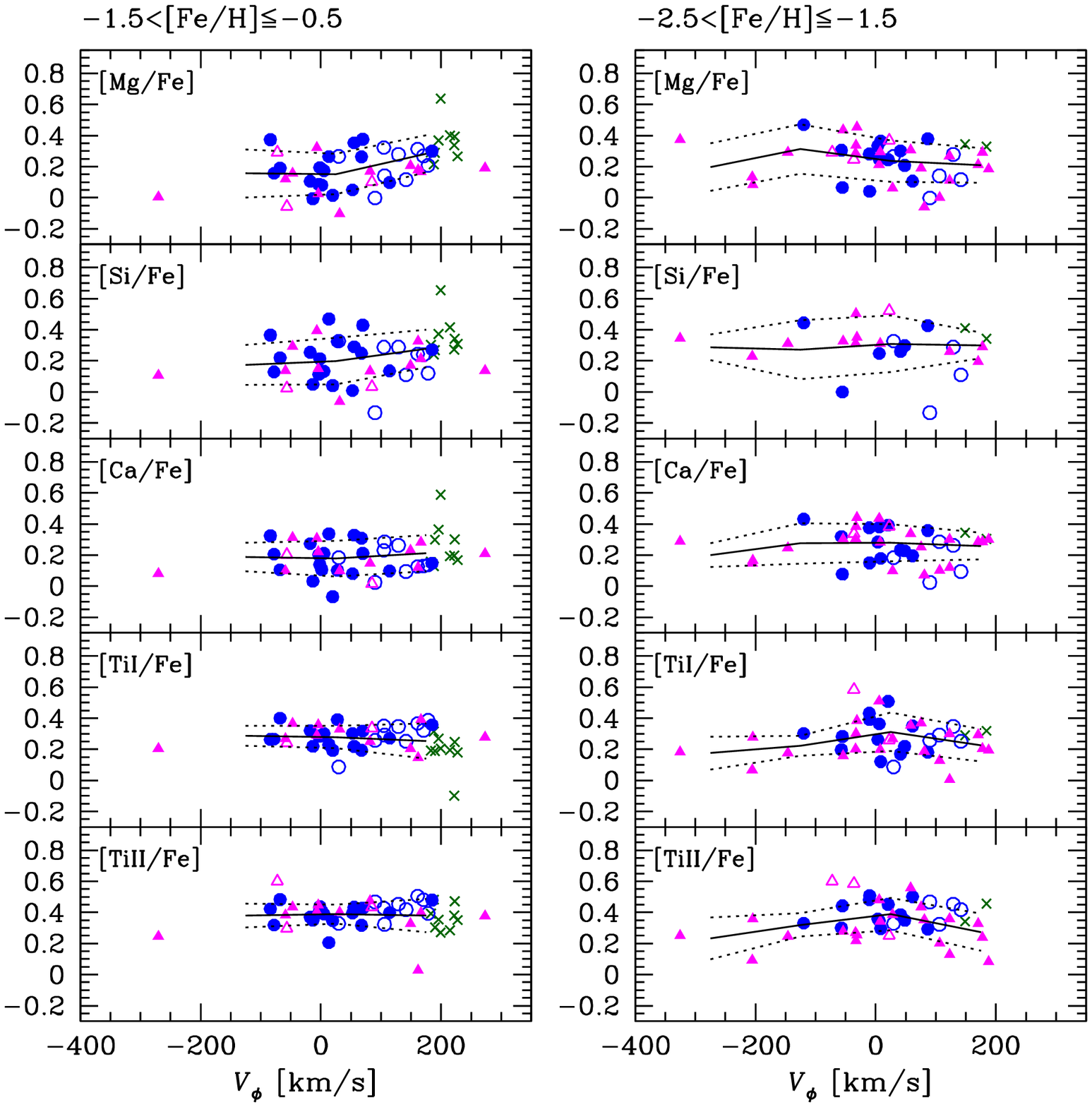}
\end{minipage}
\begin{minipage}{0.5\hsize}
\includegraphics[height=8cm]{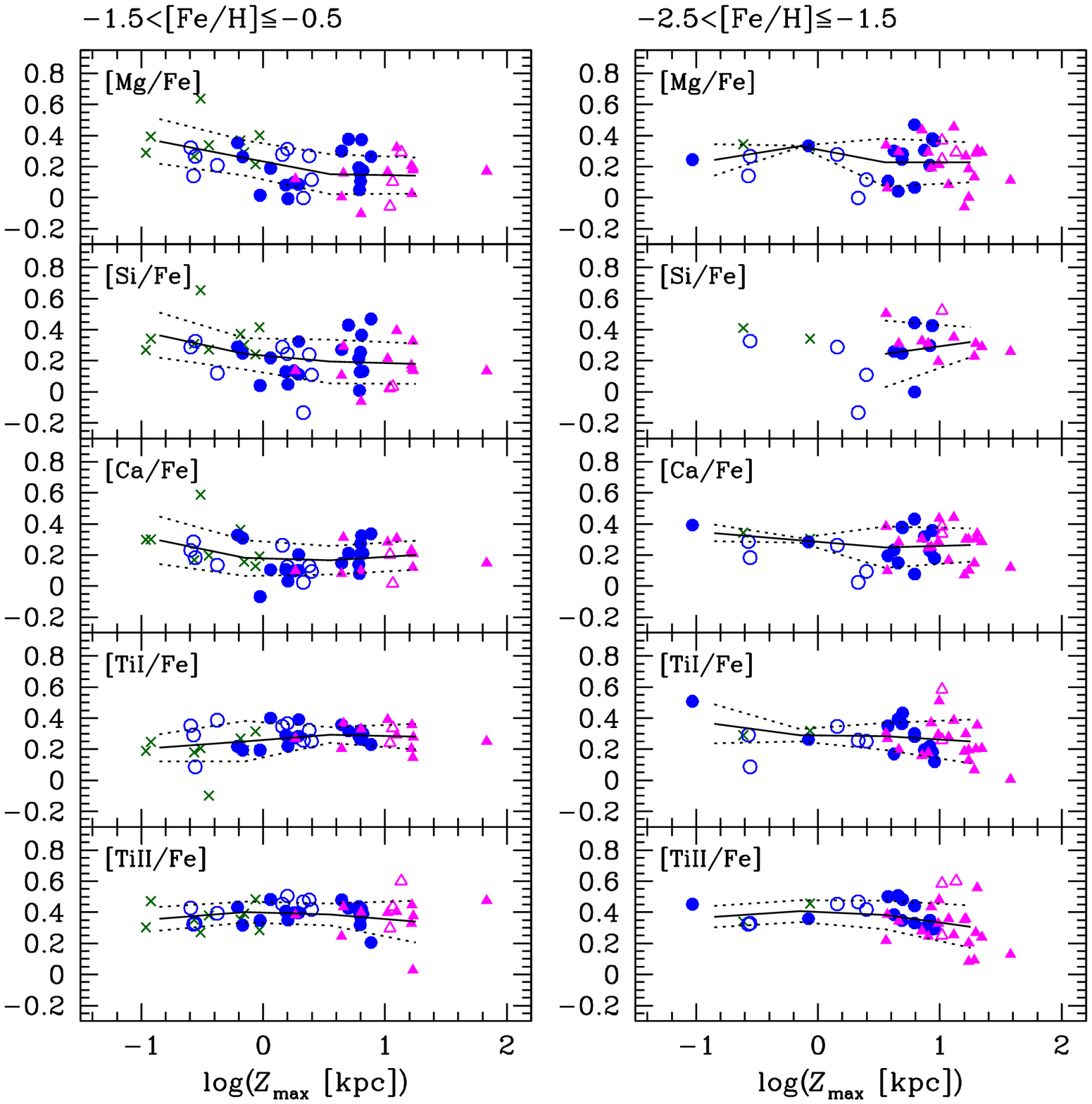}
\end{minipage}\\
\begin{minipage}{0.5\hsize}
\includegraphics[height=8cm]{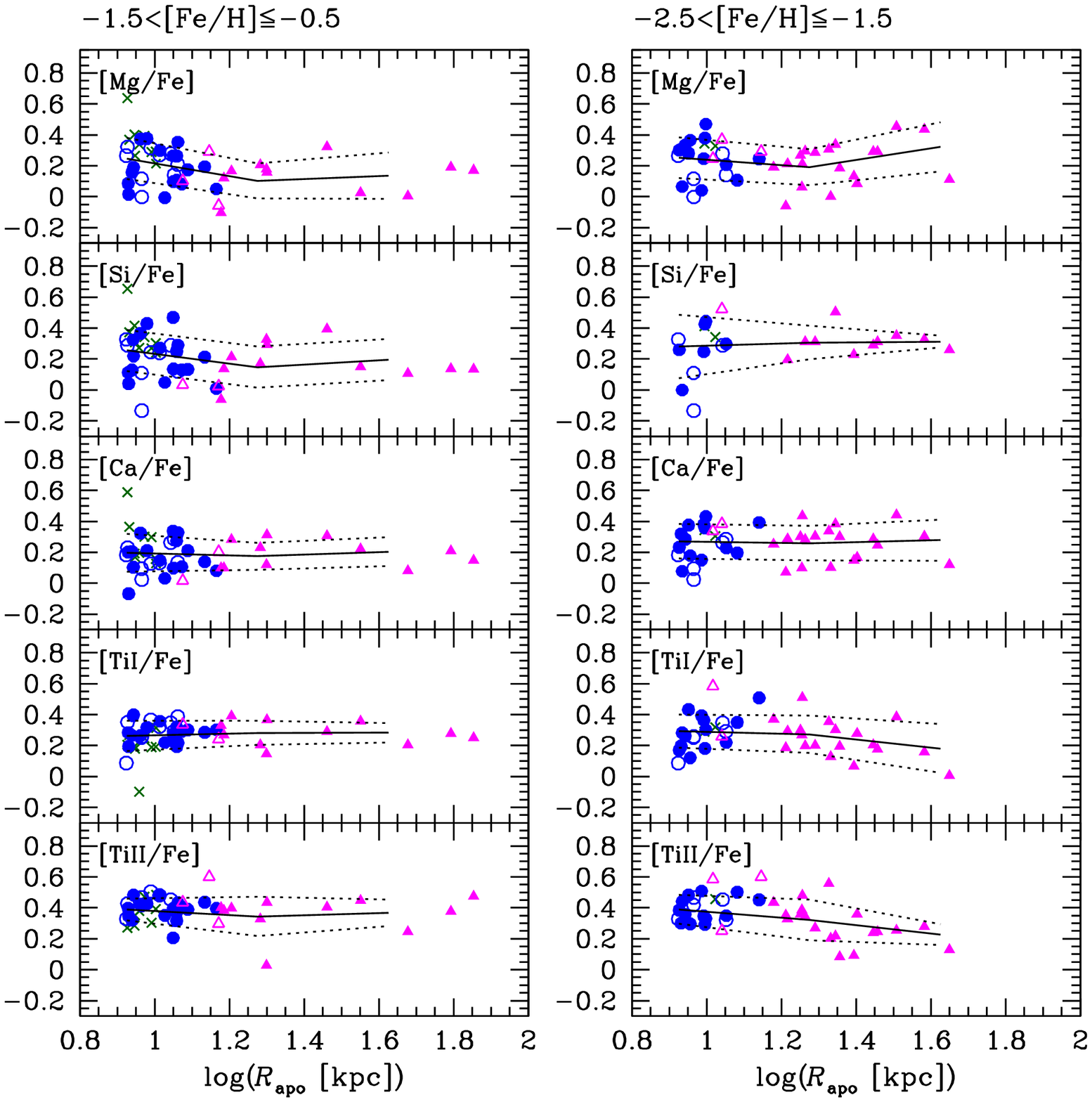}
\end{minipage}
\begin{minipage}{0.5\hsize}
\includegraphics[height=8cm]{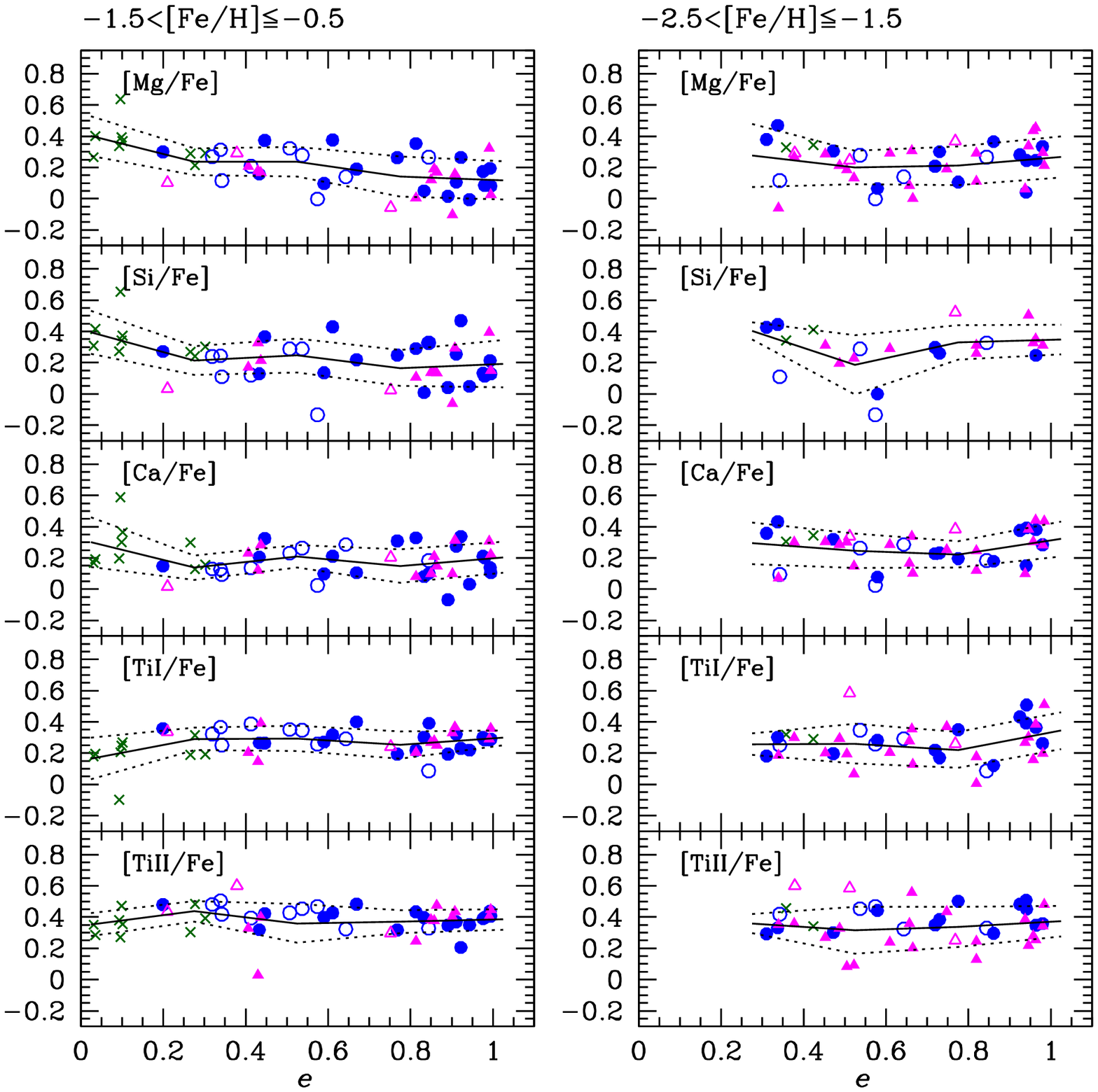}
\end{minipage}
\end{tabular}
\caption{ The [$\alpha$/Fe] ratios plotted against $V_{\phi}$ (top left), $\log Z_{\rm max}$ (top right), 
$\log R_{\rm apo}$ (bottom left) and $e$ (bottom right). 
Symbols are the same as in the top panels of Fig.\ref{fig:samplekin}. 
Solid and dashed lines 
connect means and means $\pm$ standard deviations, respectively, 
within a given interval of each orbital parameter.}
\label{fig:alpha_kin}
\end{figure}

The second possibility is that the halo stars were originally formed
within isolated dwarf galaxies that are later
accreted to the Galaxy. The accreted dwarf galaxies 
would be tidally disrupted as they orbit around the Galaxy. 
The debris stars then populated the halo while kept their
 orbital velocity similar to their disrupted host galaxies.
In this case, the stellar mass of the halo grows more 
slowly than the former case. Chemical abundance of the 
halo stars, then, would reflect metal-enrichment history 
in their progenitor dwarf galaxy, that would have less 
efficient star formation history because of their 
shallower potential well than that of the Galaxy, which makes 
the enriched gas easier to escape. Since many of the 
surviving MW dwarf satellites show lower [$\alpha$/Fe] that 
are indicative of their lower star formation rate, it is 
naively expected that the halo stars would also have lower 
[$\alpha$/Fe] ratios if the major fraction of the halo stars 
have been accreted from such a system. 
In the present study, the lower 
[Mg/Fe], [Si/Fe] or [Ca/Fe] in [Fe/H]$>-1.5$ than the 
thick disk subsample is 
found for both the inner and the outer halo subsamples.
The lower [$\alpha$/Fe] for some of the inner and outer halo stars 
may indicate that the progenitors of these stars have enriched with 
metals via Type Ia SNe and/or galactic winds are efficient 
in ejecting metals out of the system, as suggested 
for the MW dwarf satellites 
\citep{kirby11}. Then, the decreasing [Mg/Fe] with increasing 
$e$ seen in Figure \ref{fig:alpha_kin} could 
be interpreted as fraction of the accreted stars might increase with 
$e$ in the present sample.

Even though some halo star in our sample 
show relatively low [$\alpha$/Fe], that is not 
as significant as found in stars in the dwarf galaxies 
as mentioned in Section \ref{sec:dsph}. 
The difference in the [$\alpha$/Fe] ratios between the 
MW stellar halo and the dwarf satellites can be explained 
if the progenitors of the stellar halo have been accreted 
early on stopping its star formation,  
while the surviving satellites have chemically 
enriched for a longer time \citep{font06a}. 
The implication from the simulation of \citet{font06c} 
further suggests that the lack of large-scale gradients in 
the [$\alpha$/Fe] (Figure \ref{fig:alpha_kin}) 
supports the early accretions ($\gtrsim 10$ Gyr ago) 
for the MW halo. The apparent lack of the [$\alpha$/Fe] 
correlation with $\log Z_{\rm max}$ and $\log R_{\rm max}$ 
may support the early accretion scenario.  
The quiescent accretion history of 
the MW halo for the last $\sim 10$ Gyrs 
is also supported by the recent finding of 
\citet{schuster11} that their sample of metal-rich 
halo stars at [Fe/H]$\sim-1$, 
are older than $\sim 10$ Gyrs,
including relatively younger ``low-$\alpha$'' stars.

\citet{purcell10} suggest another origin of the halo stars; disk 
stars can be dynamically heated to become halo stars via minor mergers 
on the disk plane.
In the present sample, the stars classified as having 
the intermediate kinematics between the thick disk and the 
inner halo stars, shown in the open blue circles in Figure
\ref{fig:mg_si_ca}, show relatively high [Mg/Fe] ratios 
similar to the thick disk stars. This result suggests
 that the stars having moderately disk-like orbit 
could have been formed in a rapid star formation event 
like the thick disk stars. 

Cosmological simulations of \citet{zolotov10} and \citet{font11} suggest 
that a hybrid scenario for the halo formation is naturally expected 
in the hierarchical formation of the MW.
More specifically, recent hydrodynamical simulations of 
\citet{font11} suggest that, 
at a Galactocentric distance 
grater than 20 kpc, where the halo component dominates over the 
bulge component, about 20\% 
of stars may have formed in situ, while the other 
fraction of stars formed within satellite galaxies.

The possible boundary at which transition in the dominant progenitors 
(``in situ'', accreted dSph, heated disk) may occur is 
not very well constrained in the present study. 
If any, the question of whether the 
transition occurs sharply or mildly would be important 
in constraining the merging history of our Galaxy. 
More qualitative conclusion on this issue should only 
be made after the sample is significantly expanded. 
Current sample stars are all located in the solar 
neighborhood ($<2$ kpc), in which the fraction of the halo stars 
is very low. 
Obviously, larger volumes in the halo should be investigated in order to 
construct a sample representative of the halo population. 
Since high resolution spectroscopy is not efficient for 
a larger sample including distant objects beyond the 
solar neighborhood, lower resolution 
spectroscopy combined with a sufficient analysis scheme 
\citep[e.g.,][]{kirby10,lee11,ruchti11} would 
be extremely useful to explore properties of the more distant halo. 
Recently discovered substructures in the stellar halo 
would be possible candidates of recent accretion events,  
whose [$\alpha$/Fe] would give information on their progenitors.

$\alpha$-element alone may not be enough to 
constrain the chemical evolution of stellar system, since 
enrichment mechanisms other than Type II/Ia SNe are also 
thought to have played an important role.
For example, the importance of neutron-capture elements that are 
synthesized via slow-neutron capture in the asymptotic giant branch stars 
has been emphasized 
in chemical evolution models \citep[e.g.,][]{lanfranchi08}
and the observation of the dwarf satellites
showing significant Ba enhancement. Analysis of 
such neutron-capture elements as well as iron peak elements 
will be presented in the forthcoming paper 
(M. N. Ishigaki et al. in preparation). 
 
\section{Conclusion}

We have presented the abundances of Fe, Mg, Si, Ca and Ti  
for 97 metal-poor stars 
covering a wide range of metallicity ($-3.3<$[Fe/H]$<-0.5$) 
and orbital parameters, including 
those having the extreme outer halo kinematics. 
The abundances were obtained 
from the high-resolution spectra taken with Subaru/HDS using a 
one-dimensional LTE abundance analysis code with Kurucz model atmospheres. 
Our results provide insights about differences and similarities in
the [$\alpha$/Fe] ratios as a function of [Fe/H] and the kinematics 
among the three presumably old Galactic components, the thick disk, 
inner, and outer stellar halo as summarized below:

\begin{itemize}
\item The thick disk stars show high [Mg/Fe] and [Si/Fe] 
ratios than the inner and outer halo stars in their overlapping 
metallicity range ([Fe/H]$>-1.5$). The high abundance ratios 
for Mg and Si, that are predominantly synthesized in massive stars, 
are in good agreement with previous studies. This result suggests 
that the thick disk stars were formed out of gas 
primarily enriched by Type II SNe 
of massive stars with little contribution of Fe from Type Ia SNe. 
\item The inner halo stars span a wide range in [Mg/Fe] and [Si/Fe] 
ratios compared to the thick disk stars. 
The results imply that the inner halo stars have formed in 
various formation sites, presumably including the collapsed gas 
in the early Galaxy or dwarf galaxies accreted early times.    
\item The outer halo stars show similar mean [Mg/Fe] and [Si/Fe] 
ratios as the inner halo stars, which is lower than those of the 
thick disk stars. 
The inner and the outer halo stars together 
show a hint of a decreasing trend in 
[Mg/Fe], [Si/Fe] and [Ca/Fe] with [Fe/H] in [Fe/H]$>-2.0$. 
These results suggest that the contribution from Type Ia SNe 
may have played some role in the chemical evolution of the 
outer halo progenitor.
\item The three subsamples show largely overlapping distributions 
in the [Ca/Fe], [\ion{Ti}{1}/Fe] and [\ion{Ti}{2}/Fe] versus [Fe/H] diagrams. 
The different nucleosynthesis site for these two elements from 
that of Mg and Si may explain this result. However, because of 
the large errors in [\ion{Ti}{1}/Fe] and [\ion{Ti}{2}/Fe], presumably due to 
the NLTE effects, definitive conclusions from the Ti abundances 
cannot be made in the present study. 
\item Significant correlation of the [X/Fe] ratios with 
the orbital parameters ($V_{\phi}$, $Z_{\rm max}$, $R_{\rm apo}$) 
is not observed in the present sample, except for the decreasing [Mg/Fe] trend 
with the increasing $e$ in $-1.5<$[Fe/H]$\leq -0.5$. 
This result suggests that dwarf galaxies having low [X/Fe] ratios 
have not significantly contributed to build up the present-day 
stellar halo. Since this conclusion is drawn only from the local sample, 
the abundance data for a larger volume of the Galaxy are desirable 
to evaluate the large-scale abundance gradient.
\end{itemize}

More quantitative conclusions on the large-scale distribution 
of [$\alpha$/Fe] are expected by next-generation multi-object 
spectroscopic surveys that cover a larger volume of the MW thick 
disk and stellar halo.

\acknowledgments 
The authors thank the referee for her/his constructive 
comments and useful suggestions that have helped us to improve 
our paper.  
We thank A. Tajitsu, T-S. Pyo  and the staff members of 
Subaru telescope for their helpful support and assistance in our 
HDS observation. MI is grateful to U. Heiter, A. Korn 
and B. Edvardsson for the important suggestions on the 
abundance analysis and for the kind hospitality. M.I. is also 
grateful to P. E. Nissen 
for the valuable discussion and comments.
This work is supported in part from Grant-in-Aid for Scientific 
Research (23740162, 23224004) of the Ministry of Education, Culture, Sports, Science 
and Technology in Japan.



{\it Facilities:} \facility{Subaru (HDS)}.



\appendix





\clearpage



\clearpage



\clearpage







\clearpage

\begin{deluxetable}{lccccccccccc}
\tabletypesize{\scriptsize}
\tablecaption{Mean Velocities and Dispersions for the Thick disk, Inner, and Outer halo components adopted from \citet{carollo10}\label{tab:kinparams}}
\tablewidth{0pt}
\tablehead{
\colhead{Component} & \colhead{$<V_{R}>$\tablenotemark{a}} & \colhead{$<V_{\phi}>$ } & \colhead{$<V_{Z}>$ } & \colhead{$\sigma_{R}$\tablenotemark{b}} &
\colhead{$\sigma_{\phi}$} & \colhead{$\sigma_{Z}$} & \multicolumn{5}{c}{$f_{i}$ ($Z_{\rm max}$ kpc)\tablenotemark{c}} \\ \cline{8-12}
\colhead{} & \colhead{(km s$^{-1}$)} & \colhead{(km s$^{-1}$)} & \colhead{(km s$^{-1}$)} & \colhead{(km s$^{-1}$)} &
\colhead{(km s$^{-1}$)} & \colhead{(km s$^{-1}$)} & \colhead{$<5$} & \colhead{$5-10$} & \colhead{$10-15$}& \colhead{$15-20$} & \colhead{$>20$}
}
\startdata
Thick disk & 3 & 182 & 0 & 53 & 51 & 35 &0.55 & 0.00& 0.00& 0.00& 0.00\\
Inner halo & 3 & 7 & 3& 150& 95 & 85 & 0.45 & 1.00 & 0.80 & 0.55 & 0.08\\
Outer halo & $-$9& $-$80 &2 &159 & 165 & 116 & 0.00 & 0.00 & 0.20 & 0.45 & 0.92\\ 
\enddata
\tablecomments{}
\tablenotetext{a}{Mean velocities in a cylindrical coordinate.}
\tablenotetext{b}{Velocity dispersions in a cylindrical coordinate.}
\tablenotetext{c}{Fractional contribution of stars at a given $Z_{\rm max}$ range.}
\end{deluxetable}


\clearpage

\begin{deluxetable}{lcccccccc}
\tabletypesize{\scriptsize}
\tablecaption{ Kinematics of the sample \label{tab:kinematics}}
\tablewidth{0pt}
\tablehead{
\colhead{Object name} & \colhead{[Fe/H]\tablenotemark{a}} & \colhead{$V_{\rm rad}$\tablenotemark{b}} & \colhead{$V_{R}$} & \colhead{$V_{\phi}$} &\colhead{$V_{Z}$} & \colhead{$P_{\rm TD}$\tablenotemark{c}} & \colhead{$P_{\rm IH}$} & \colhead{$P_{\rm OH}$} \\
\colhead{ } & \colhead{(dex)} & \colhead{(km s$^{-1}$)} & \colhead{(km s$^{-1}$)} & \colhead{(km s$^{-1}$)} &\colhead{(km s$^{-1}$)} & \colhead{ } & \colhead{ } & \colhead{} 
}
\startdata

BD+01\arcdeg 3070 &   $-$1.37 &  $-$328.4 $\pm$  0.3 &     338.9 $\pm$      24.1 &     273.0 $\pm$      11.1 &    $-$113.2 $\pm$      22.8 &    0.00 &    0.00 &    1.00\\
BD--03\arcdeg 5215 &   $-$1.48 &  $-$296.5 $\pm$  0.3 &     147.5 $\pm$       1.8 &      13.2 $\pm$       7.4 &     143.1 $\pm$       7.8 &    0.00 &    1.00 &    0.00\\
BD+04\arcdeg 2466 &   $-$1.95 &    33.4 $\pm$  0.5 &     $-$56.6 $\pm$       8.7 &    $-$119.7 $\pm$      68.3 &    $-$151.1 $\pm$      37.6 &    0.00 &    1.00 &    0.00\\
BD+04\arcdeg 2621 &   $-$2.39 &   $-$40.0 $\pm$  0.3 &      $-$1.7 $\pm$      10.3 &     $-$57.0 $\pm$      61.4 &    $-$156.5 $\pm$      25.6 &    0.00 &    1.00 &    0.00\\
BD--08\arcdeg 3901 &   $-$1.56 &  $-$108.9 $\pm$  0.3 &     109.6 $\pm$       8.5 &     184.3 $\pm$      13.1 &     $-$31.3 $\pm$       7.1 &    0.92 &    0.08 &    0.00\\
\enddata
\tablecomments{Table \ref{tab:kinematics} is published in its entirety in the electronic 
edition of the Astrophysical Journal. A portion is shown here for guidance regarding its form and context.}
\tablenotetext{a}{[Fe/H] estimated from the high-resolution spectra obtained with Subaru/HDS.}
\tablenotetext{b}{Radial velocities measured from the high-resolution spectra obtained with Subaru/HDS.}
\tablenotetext{c}{See Section 2.2.2.}
\end{deluxetable}

\clearpage

\begin{deluxetable}{lccccccc}
\tabletypesize{\tiny}
\tablecaption{ Summary of the new observation \label{tab:obssummary}}
\tablewidth{0pt}
\tablehead{
\colhead{Object name} & \colhead{RA} & \colhead{DEC} & \colhead{$V$} & \colhead{Date} &
\colhead{Exp.time } & \colhead{N\tablenotemark{a}} & \colhead{S/N\tablenotemark{b}} \\
\colhead{} & \colhead{} & \colhead{} & \colhead{(mag)} & \colhead{} &
\colhead{(s)} & \colhead{} & \colhead{} 
}
\startdata
     G176--53 &   11:46:34.100 &   +50:52:36.52 &   9.92 &   2010-05-26 &  900 &  1 &  269 \\
    HD 103295 &   11:53:37.900 &   $-$28:38:01.05 &   9.58 &   2010-05-26 &  900 &  1 &  291 \\
    HD 233891 &   11:59:58.915 &   +51:46:05.04 &   8.80 &   2010-05-26 &  400 &  1 &  280 \\
    HD 105004 &   12:05:25.548 &   $-$26:35:36.09 &  10.31 &   2010-05-26 &  600 &  2 &  180 \\
       G59--1 &   12:08:54.954 &   +21:47:11.97 &   9.52 &   2010-05-26 &  900 &  1 &  302 \\
    HD 105546 &   12:09:02.284 &   +59:00:50.97 &   8.61 &   2010-05-26 &  400 &  1 &  302 \\
    HD 106373 &   12:14:14.150 &   $-$28:14:56.36 &   8.91 &   2010-05-26 &  600 &  1 &  349 \\
    HD 108317 &   12:26:37.563 &   +05:18:08.16 &   8.03 &   2010-05-26 &  200 &  2 &  299 \\
    HD 108976 &   12:31:03.580 &   +27:43:40.09 &   8.56 &   2010-05-26 &  400 &  1 &  353 \\
    HD 109995 &   12:38:47.757 &   +39:18:18.84 &   7.60 &   2010-05-26 &  200 &  1 &  383 \\
      G60--26 &   12:40:00.196 &   +12:38:26.17 &   9.82 &   2010-05-26 &  900 &  1 &  284 \\
    HD 112126 &   12:53:49.591 &   +32:30:17.23 &   8.74 &   2010-05-26 &  400 &  1 &  279 \\
      G14--33 &   13:08:48.852 &   $-$03:58:09.48 &  11.18 &   2010-05-26 & 1200 &  2 &  177 \\
    HD 116064 &   13:21:44.254 &   $-$39:18:29.20 &   8.81 &   2010-05-26 &  600 &  1 &  332 \\
BD+09\arcdeg2776 &   13:33:32.853 &   +08:35:09.18 &   7.96 &   2010-05-26 &  300 &  1 &  360 \\
      G63--46 &   13:40:00.438 &   +12:35:13.22 &   9.39 &   2010-05-26 &  600 &  1 &  280 \\
    HD 122196 &   14:01:02.773 &   $-$38:02:52.04 &   8.72 &   2010-05-26 &  415 &  1 &  283 \\
    HD 122956 &   14:05:13.431 &   $-$14:51:09.25 &   7.22 &   2010-05-26 &  100 &  1 &  288 \\
      G66--51 &   15:00:50.751 &   +02:07:35.50 &  10.63 &   2010-05-26 &  600 &  2 &  161 \\
BD-08\arcdeg3901 &   15:04:52.065 &   $-$08:48:56.60 &   9.47 &   2010-05-26 &  600 &  1 &  261 \\
     G153--21 &   16:03:00.868 &   $-$06:27:06.83 &  10.19 &   2010-05-26 &  600 &  2 &  196 \\
    HD 171496 &   18:36:08.100 &   $-$24:25:55.66 &   8.49 &   2010-05-26 &  400 &  1 &  299 \\
    LP 751-19 &   18:51:09.320 &   $-$11:48:06.99 &  10.42 &   2010-05-26 &  600 &  2 &  156 \\
    LTT 15637 &   19:15:07.514 &   +10:34:46.43 &   9.42 &   2010-05-26 &  600 &  1 &  272 \\
    HD 184266 &   19:34:15.916 &   $-$16:18:44.80 &   7.59 &   2010-05-26 &  200 &  1 &  331 \\
     G142--44 &   19:38:53.270 &   +16:25:52.01 &  11.15 &   2010-06-18 & 1000 &  2 &  155 \\
      G23--14 &   19:51:49.475 &   +05:37:01.20 &  10.71 &   2010-06-18 &  800 &  2 &  173 \\
    HD 188510 &   19:55:09.157 &   +10:44:44.85 &   8.83 &   2010-05-26 &  400 &  1 &  250 \\
       G24--3 &   20:05:43.948 &   +04:03:05.98 &  10.46 &   2010-05-26 &  600 &  2 &  153 \\
    HD 193901 &   20:23:36.569 &   $-$21:22:08.94 &   8.66 &   2010-05-26 &  400 &  1 &  254 \\
    HD 196892 &   20:40:49.665 &   $-$18:47:21.00 &   8.25 &   2010-05-26 &  400 &  1 &  347 \\
     G210--33 &   20:45:21.964 &   +40:23:19.68 &  11.20 &   2010-06-18 &  900 &  2 &  153 \\
BD-14\arcdeg5850 &   20:47:35.059 &   $-$14:25:35.90 &  10.96 &   2010-06-18 &  700 &  2 &  148 \\
      G212--7 &   20:55:15.410 &   +42:17:54.06 &  10.27 &   2010-05-26 &  600 &  2 &  174 \\
    HD 199854 &   21:00:13.794 &   $-$15:06:35.32 &   8.95 &   2010-05-26 &  600 &  1 &  309 \\
     G187--40 &   21:21:56.700 &   +27:27:12.45 &  10.51 &   2010-06-18 &  600 &  2 &  164 \\
BD+46\arcdeg3330 &   21:28:47.483 &   +47:06:53.65 &   9.30 &   2010-05-26 &  600 &  1 &  325 \\
     G231--52 &   21:39:13.871 &   +60:16:56.82 &  10.34 &   2010-06-18 &  500 &  2 &  149 \\
     G188--22 &   21:43:56.002 &   +27:23:33.22 &  10.05 &   2010-05-26 &  600 &  2 &  193 \\
BD+47\arcdeg3617 &   21:57:01.410 &   +48:22:46.66 &  10.30 &   2010-06-18 &  500 &  2 &  214 \\
BD+46\arcdeg3563 &   22:04:13.870 &   +47:24:13.38 &  10.10 &   2010-06-18 &  900 &  1 &  292 \\
    HD 210295 &   22:09:41.566 &   $-$13:36:03.37 &   9.57 &   2010-05-26 &  900 &  1 &  260 \\
    HD 213487 &   22:32:03.550 &   $-$21:35:40.04 &   9.87 &   2010-05-26 &  900 &  1 &  235 \\
    HD 213467 &   22:32:08.605 &   $-$31:10:10.22 &   8.52 &   2010-05-26 &  400 &  1 &  286 \\
    HD 215601 &   22:46:48.598 &   $-$31:52:04.73 &   8.46 &   2010-05-26 &  400 &  1 &  311 \\
\enddata
\tablenotetext{a}{Number of exposures.}
\tablenotetext{b}{Signal-nose-ratio per resolution element.}
\end{deluxetable}

\clearpage

\begin{deluxetable}{lcccccccc}
\tabletypesize{\scriptsize}
\tablecaption{ Equivalent widths \label{tab:ews}}
\tablewidth{0pt}
\tablehead{
\colhead{Object name} & \colhead{Z/Ion} & \colhead{Element} & \colhead{$\lambda$} & \colhead{$\log gf$} &
\colhead{$\chi$} & \colhead{EW} & \colhead{Flag\tablenotemark{a}} &\colhead{Refs.\tablenotemark{b}} \\
\colhead{} & \colhead{} & \colhead{} & \colhead{({\AA})} & \colhead{(dex)} &
\colhead{(eV)} & \colhead{({\AA})} & \colhead{} & \colhead{} 
}
\startdata
BD+01\arcdeg3070 & 26 1 &   FeI &    4114.44 &  $-$1.30 &   2.83 &   55.69 &  1 & O91            \\
BD+01\arcdeg3070 & 26 1 &   FeI &    4132.90 &  $-$1.01 &   2.84 &   66.03 &  1 & O91            \\
BD+01\arcdeg3070 & 26 1 &   FeI &    4147.67 &  $-$2.10 &   1.48 &   79.11 &  1 & B80            \\
BD+01\arcdeg3070 & 26 1 &   FeI &    4184.89 &  $-$0.87 &   2.83 &   70.24 &  1 & O91            \\
BD+01\arcdeg3070 & 26 1 &   FeI &    4222.21 &  $-$0.97 &   2.45 &   84.88 &  0 & B82a           \\
\enddata
\tablecomments{Table \ref{tab:ews} is published in its entirety in the electronic 
edition of the Astrophysical Journal. A portion is shown here for guidance regarding its form and context.}
\tablenotetext{a}{1: Used in the abundance analysis, 0: Not used in the abundance analysis.}
\tablenotetext{b}{Reference of adopted $\log gf$. A complete list of references 
are given in the electronic version of this table.}
\end{deluxetable}

\clearpage

\begin{deluxetable}{lcccccccccc}
\rotate
\tabletypesize{\scriptsize}
\tablecaption{ Atmospheric parameters and abundances \label{tab:stpm_ab}}
\tablewidth{0pt}
\tablehead{
\colhead{Object name} & \colhead{$T_{\rm eff}$} & \colhead{$\log g$} & \colhead{$\xi$} & \colhead{[Fe I/H]} & \colhead{[Fe II/H]} & \colhead{[Mg/Fe]} &\colhead{[Si/Fe]} & \colhead{[Ca/Fe]} & \colhead{[Ti I/Fe]} & \colhead{[Ti II/Fe]}\\
\colhead{} & \colhead{(K)} & \colhead{(dex)} & \colhead{(km s$^{-1}$)} & \colhead{(dex)} & \colhead{(dex)} & \colhead{(dex)} &\colhead{(dex)} & \colhead{(dex)} & \colhead{(dex)} & \colhead{(dex)}

}
\startdata

BD+01\arcdeg3070 & 5404 &    3.65 &    1.18 &   $-$1.38 $\pm$   0.14 &   $-$1.36 $\pm$   0.13 &    0.19 $\pm$   0.11 &    0.14 $\pm$   0.15 &    0.21 $\pm$   0.08 &    0.28 $\pm$   0.08 &    0.38 $\pm$   0.08 \\
BD+04\arcdeg2466 & 5223 &    2.02 &    1.72 &   $-$1.95 $\pm$   0.14 &   $-$1.94 $\pm$   0.12 &    0.47 $\pm$   0.10 &    0.44 $\pm$   0.12 &    0.43 $\pm$   0.08 &    0.30 $\pm$   0.10 &    0.33 $\pm$   0.11 \\
BD+04\arcdeg2621 & 4754 &    1.63 &    1.72 &   $-$2.37 $\pm$   0.16 &   $-$2.41 $\pm$   0.12 &    0.30 $\pm$   0.09 &   $-$9.99 $\pm$   0.00 &    0.32 $\pm$   0.09 &    0.20 $\pm$   0.08 &    0.30 $\pm$   0.09 \\
BD+09\arcdeg2870 & 4632 &    1.30 &    1.63 &   $-$2.37 $\pm$   0.17 &   $-$2.41 $\pm$   0.12 &    0.43 $\pm$   0.09 &    0.33 $\pm$   0.14 &    0.30 $\pm$   0.09 &    0.16 $\pm$   0.08 &    0.28 $\pm$   0.09 \\
BD+10\arcdeg2495 & 4973 &    2.25 &    1.64 &   $-$2.02 $\pm$   0.15 &   $-$2.02 $\pm$   0.12 &    0.29 $\pm$   0.09 &    0.29 $\pm$   0.14 &    0.29 $\pm$   0.08 &    0.20 $\pm$   0.08 &    0.24 $\pm$   0.07 \\
\enddata
\tablecomments{Table \ref{tab:stpm_ab} is published in its 
entirety in the electronic edition of the Astrophysical Journal. 
A portion is shown here for guidance regarding its form and context.}
\end{deluxetable}

\clearpage

\begin{deluxetable}{lccccccccccccc}
\rotate
\tabletypesize{\scriptsize}
\tablecaption{ Comparison with NS10 \label{tab:comp_ns10}}
\tablewidth{-1pt}
\tablehead{
\colhead{Starname}&\colhead{NS10/TW} & \colhead{$T_{\rm eff}$} & \colhead{$\log g$} & \colhead{$\xi$} & \colhead{[Fe/H] } &
\colhead{[Mg/Fe]} & \colhead{[Si/Fe]} & \colhead{[Ca/Fe]} &\colhead{[Ti/Fe]} & \colhead{$U$\tablenotemark{a}}& \colhead{$V$\tablenotemark{a}} & \colhead{$W$\tablenotemark{a}} & \colhead{Classification}\\
\colhead{} & \colhead{}&\colhead{(K)} & \colhead{(dex)} & \colhead{(km s$^{-1}$)} & \colhead{(dex)} &\colhead{(dex)} & \colhead{(dex)} & \colhead{(dex)} &\colhead{(dex)} & \colhead{(km s$^{-1}$)}& \colhead{(km s$^{-1}$)} & \colhead{(km s$^{-1}$)} & \colhead{}
}
\startdata
     G 112--43    &NS10& 6074&   4.03&   1.30& $-$1.25&  0.21&  0.15&  0.29&  0.29&  $-$145&  $-$119&  $-$293&    low-alpha\\
                & TW & 6176&   4.05&  1.36&   $-$1.33&   0.17&   0.21&   0.28&   0.39&  $-$126&   $-$52&  $-$197 & OH\\
     G 53--41     &NS10& 5859&   4.27&   1.30& $-$1.20&  0.24&  0.24&  0.31&  0.14&    31&  $-$299&  $-$150&    low-alpha\\
                & TW & 6070&   4.56&  0.77&   $-$1.15&   0.16&   0.13&   0.21&   0.27&    27&  $-$298&  $-$153 & IH\\
     G 125--13    &NS10& 5848&   4.28&   1.50& $-$1.43&  0.30&  0.28&  0.30&  0.20&  $-$215&  $-$228&  $-$157& (high-alpha)\\
                & TW & 6079&   4.75&  0.79&   $-$1.35&   0.17&   0.13&   0.21&   0.30&  $-$170&  $-$217&  $-$130 & IH\\
     HD111980   &NS10& 5778&   3.96&   1.50& $-$1.08&  0.36&  0.40&  0.34&  0.25&  $-$239&  $-$174&   $-$57&   high-alpha\\
                & TW & 5798&   4.04&  1.21&   $-$1.13&   0.32&   0.39&   0.31&   0.29&  $-$327&  $-$224&  $-$131 & OH\\
     G 20--15     &NS10& 6027&   4.32&   1.60& $-$1.49&  0.22&  0.23&  0.29&  0.24&  $-$161&   $-$60&  $-$210&  (low-alpha)\\
                & TW & 6042&   4.26&  1.22&   $-$1.62&   0.21&   0.19&   0.28&   0.29&  $-$150&   $-$50&  $-$183 & OH\\
     HD105004   &NS10& 5754&   4.30&   1.20& $-$0.82&  0.17&  0.14&  0.17&  0.07&    44&  $-$239&   $-$92&    low-alpha\\
                & TW & 6115&   5.00&  0.40&   $-$0.60&   0.02&   0.04&  $-$0.07&   0.19&    21&  $-$200&   $-$52 & IH\\
     G 176--53    &NS10& 5523&   4.48&   1.00& $-$1.34&  0.15&  0.15&  0.25&  0.15&   230&  $-$271&    61&    low-alpha\\
                & TW & 5753&   5.00&  0.17&   $-$1.28&   0.12&   0.14&   0.10&   0.27&   232&  $-$279&    60 & OH\\
     HD193901   &NS10& 5650&   4.36&   1.20& $-$1.09&  0.13&  0.18&  0.22&  0.10&   148&  $-$233&   $-$66&    low-alpha\\
                & TW & 5908&   4.94&  0.29&   $-$0.93&  $-$0.01&   0.05&   0.03&   0.22&   146&  $-$233&   $-$65 & IH\\
     G 188--22    &NS10& 5974&   4.18&   1.50& $-$1.32&  0.39&  0.37&  0.37&  0.28&  $-$193&   $-$99&    71&   high-alpha\\
                & TW & 6170&   4.52&  1.13&   $-$1.28&   0.28&   0.29&   0.26&   0.35&  $-$135&   $-$93&    62 & IH/TD\\
\enddata
\tablenotetext{a}{Adopted solar motions are $(U_{\sun}, V_{\sun}, W_{\sun})=(-7.5,13.5,6.8)$ km s$^{-1}$ in NS10 and $(-9,12,7)$ km s$^{-1}$ in TW.}
\end{deluxetable}

\begin{deluxetable}{lccccccccc}
\tabletypesize{\scriptsize}
\tablecaption{ Error estimates \label{tab:error}}
\tablewidth{-1pt}
\tablehead{
\colhead{Object name} & \colhead{$T_{\rm eff}$} & \colhead{$\log g$} & \colhead{$\xi$} & \colhead{Element} &
\colhead{Abundance} & \colhead{$\sigma_{T_{\rm eff}}\pm 100$} & \colhead{$\sigma_{\log g}\pm 0.3$} &\colhead{$\sigma_{\xi}\pm 0.3$} & \colhead{MARCS}\\
\colhead{ } & \colhead{(K)} & \colhead{(dex)} & \colhead{(km s$^{-1}$)} & \colhead{ } &
\colhead{(dex)} & \colhead{(dex)} & \colhead{(dex)} &\colhead{(dex)} & \colhead{(dex)}
}
\startdata
          G24--3 &  6180 &   4.6 &   0.3 &     [Fe I/H] &   $-$1.40 &    0.08 &  $-$0.06&  $-$0.03&  $-$0.03 \\
  &   &   &   &   &   &   $-$0.09 &   0.06&  \nodata& \\
  &   &   &   &    [Fe II/H] &   $-$1.39 &    0.02 &   0.08&  $-$0.03&  $-$0.04\\
  &   &   &   &   &   &   $-$0.02 &  $-$0.08&  \nodata&\\
  &   &   &   &      [Mg/Fe] &    0.09 &   $-$0.03 &   0.01&   0.02&   0.06\\
  &   &   &   &   &   &    0.03 &  $-$0.00&  \nodata&\\
  &   &   &   &      [Si/Fe] &    0.11 &   $-$0.06 &   0.08&   0.03&   0.01\\
  &   &   &   &   &   &    0.06 &  $-$0.07&  \nodata&\\
  &   &   &   &      [Ca/Fe] &    0.20 &   $-$0.03 &   0.01&   0.02&   0.00\\
  &   &   &   &   &   &    0.03 &  $-$0.01&  \nodata&\\
  &   &   &   &    [Ti I/Fe] &    0.28 &    0.00 &   0.05&   0.01&   0.01\\
  &   &   &   &   &   &   $-$0.00 &  $-$0.05&  \nodata&\\
  &   &   &   &   [Ti II/Fe] &    0.40 &    0.03 &  $-$0.01&   0.00&  $-$0.03\\
  &   &   &   &   &   &   $-$0.03 &   0.01&  \nodata&\\
         HD215601 &  4892 &   1.6 &   1.7 &     [Fe I/H] &   $-$1.40 &    0.14 &  $-$0.02&  $-$0.10&  $-$0.16 \\
  &   &   &   &   &   &   $-$0.14 &   0.03&   0.13& \\
  &   &   &   &    [Fe II/H] &   $-$1.44 &   $-$0.03 &   0.12&  $-$0.07&  $-$0.06\\
  &   &   &   &   &   &    0.03 &  $-$0.12&   0.09&\\
  &   &   &   &      [Mg/Fe] &    0.27 &   $-$0.07 &   0.00&   0.07&   0.06\\
  &   &   &   &   &   &    0.06 &  $-$0.00&  $-$0.08&\\
  &   &   &   &      [Si/Fe] &    0.33 &   $-$0.10 &   0.03&   0.09&   0.10\\
  &   &   &   &   &   &    0.10 &  $-$0.03&  $-$0.12&\\
  &   &   &   &      [Ca/Fe] &    0.18 &   $-$0.04 &  $-$0.01&   0.02&   0.05\\
  &   &   &   &   &   &    0.04 &   0.00&  $-$0.04&\\
  &   &   &   &    [Ti I/Fe] &    0.09 &    0.01 &   0.00&   0.05&   0.07\\
  &   &   &   &   &   &   $-$0.01 &   0.00&  $-$0.06&\\
  &   &   &   &   [Ti II/Fe] &    0.33 &    0.04 &  $-$0.01&  $-$0.10&  $-$0.07\\
  &   &   &   &   &   &   $-$0.04 &   0.01&   0.12&\\
          G64--37 &  6621 &   4.6 &   2.5 &     [Fe I/H] &   $-$3.07 &    0.07 &   0.00&  $-$0.01&   0.01 \\
  &   &   &   &   &   &   $-$0.07 &   0.00&   0.01& \\
  &   &   &   &    [Fe II/H] &   $-$3.07 &    0.01 &   0.10&  $-$0.00&   0.00\\
  &   &   &   &   &   &   $-$0.01 &  $-$0.10&   0.00&\\
  &   &   &   &      [Mg/Fe] &    0.15 &   $-$0.01 &  $-$0.02&  $-$0.02&   0.00\\
  &   &   &   &   &   &    0.01 &   0.01&   0.03&\\
  &   &   &   &      [Ca/Fe] &    0.43 &   $-$0.02 &  $-$0.00&   0.00&  $-$0.01\\
  &   &   &   &   &   &    0.02 &   0.00&  $-$0.00&\\
  &   &   &   &    [Ti I/Fe] &    0.83 &    0.01 &  $-$0.00&   0.00&  $-$0.00\\
  &   &   &   &   &   &   $-$0.01 &   0.00&  $-$0.00&\\
  &   &   &   &   [Ti II/Fe] &    0.51 &    0.02 &  $-$0.00&  $-$0.00&   0.01\\
  &   &   &   &   &   &   $-$0.03 &   0.00&   0.00&\\
 BD--18\arcdeg271 &  4234 &   0.4 &   2.5 &     [Fe I/H] &   $-$2.58 &    0.16 &  $-$0.05&  $-$0.04&  \nodata \\
  &   &   &   &   &   &   $-$0.17 &   0.06&   0.05& \\
  &   &   &   &    [Fe II/H] &   $-$2.54 &   $-$0.03 &   0.11&  $-$0.01&  \nodata\\
  &   &   &   &   &   &    0.04 &  $-$0.10&   0.01&\\
  &   &   &   &      [Mg/Fe] &    0.48 &   $-$0.07 &   0.01&   0.03&  \nodata\\
  &   &   &   &   &   &    0.07 &  $-$0.01&  $-$0.04&\\
  &   &   &   &      [Si/Fe] &    0.43 &   $-$0.12 &   0.05&   0.04&  \nodata\\
  &   &   &   &   &   &    0.14 &  $-$0.05&  $-$0.05&\\
  &   &   &   &      [Ca/Fe] &    0.28 &   $-$0.06 &  $-$0.00&   0.01&  \nodata\\
  &   &   &   &   &   &    0.05 &  $-$0.00&  $-$0.02&\\
  &   &   &   &    [Ti I/Fe] &    0.19 &    0.06 &   0.00&   0.01&  \nodata\\
  &   &   &   &   &   &   $-$0.06 &  $-$0.00&  $-$0.02&\\
  &   &   &   &   [Ti II/Fe] &    0.24 &    0.03 &  $-$0.02&  $-$0.04&  \nodata\\
  &   &   &   &   &   &   $-$0.03 &   0.02&   0.05&\\
\enddata
\end{deluxetable}

\clearpage

\begin{deluxetable}{lcccccccccc}
\tabletypesize{\scriptsize}
\tablecaption{ Means and standard deviations in the abundance ratios \label{tab:mean_dev}}
\tablewidth{0pt}
\tablehead{
\colhead{[X/Fe] } & \colhead{TD/IH/OH}&\multicolumn{3}{c}{[Fe/H]$>-1.5$} & \multicolumn{3}{c}{$-2.5<$[Fe/H]$\leq-1.5$} & \multicolumn{3}{c}{[Fe/H]$\leq -2.5$} \\ \cline{3-11}
\colhead{} &\colhead{} & \colhead{$\mu$\tablenotemark{a}}& \colhead{$\sigma$\tablenotemark{b}} & \colhead{N\tablenotemark{c}} &  \colhead{$\mu$}& \colhead{$\sigma$} & \colhead{N} &  \colhead{$\mu$} & \colhead{$\sigma$} & \colhead{N} 
}
\startdata
        Mg & TD &    0.32 &    0.07 &  8 &    0.34 &    0.01 &  2 &    0.48 &    0.00 &  1 \\
           & IH &    0.18 &    0.12 & 17 &    0.26 &    0.13 & 13 &    0.16 &    0.08 &  3 \\
           & OH &    0.13 &    0.12 & 11 &    0.22 &    0.14 & 20 &    0.27 &    0.14 &  6 \\
        Si & TD &    0.32 &    0.06 &  8 &    0.38 &    0.05 &  2 &  \multicolumn{3}{c}{\nodata} \\
           & IH &    0.21 &    0.13 & 18 &    0.28 &    0.16 &  6 &  \multicolumn{3}{c}{\nodata} \\
           & OH &    0.18 &    0.12 & 11 &    0.31 &    0.08 & 11 &    0.43 &    0.00 &  1 \\
        Ca & TD &    0.23 &    0.08 &  8 &    0.33 &    0.03 &  2 &    0.41 &    0.00 &  1 \\
           & IH &    0.17 &    0.11 & 18 &    0.28 &    0.11 & 13 &    0.28 &    0.17 &  3 \\
           & OH &    0.19 &    0.09 & 11 &    0.26 &    0.11 & 20 &    0.38 &    0.13 &  5 \\
      Ti I & TD &    0.19 &    0.12 &  9 &    0.30 &    0.02 &  2 &    0.63 &    0.00 &  1 \\
           & IH &    0.28 &    0.06 & 18 &    0.29 &    0.11 & 13 &    0.41 &    0.37 &  3 \\
           & OH &    0.28 &    0.08 & 11 &    0.24 &    0.12 & 20 &    0.20 &    0.01 &  2 \\
     Ti II & TD &    0.37 &    0.08 &  9 &    0.40 &    0.08 &  2 &    0.38 &    0.00 &  1 \\
           & IH &    0.39 &    0.07 & 17 &    0.39 &    0.08 & 13 &    0.42 &    0.21 &  3 \\
           & OH &    0.36 &    0.12 & 11 &    0.29 &    0.12 & 20 &    0.50 &    0.21 &  6 \\
\enddata
\tablenotetext{a}{Means of the abundance ratios within a given [Fe/H] interval.}
\tablenotetext{b}{Standard deviations of the means.}
\tablenotetext{c}{The number of stars used to calculate the $\mu$ and $\sigma$.}
\end{deluxetable}




\end{document}